\documentclass[journal=pasa]{cup-journal}
\usepackage[british]{babel}
\usepackage{caption}
\usepackage{placeins}    
\usepackage{url}

\usepackage{subcaption}
\usepackage{graphicx}
\usepackage{adjustbox} 
\usepackage{soul}

\usepackage{longtable}
\usepackage{booktabs}
\usepackage{afterpage}

\usepackage{xcolor}
\usepackage{amsmath}
\usepackage{amssymb,microtype,siunitx,booktabs}
\usepackage[colorlinks=true, linkcolor=blue, citecolor=blue, urlcolor=blue]{hyperref}

\title{Radio Monitoring of Classical Novae using the ASKAP Variable and Slow Transients Survey}

\author{Aishani Majumder}
\author{David L.~Kaplan}
\affiliation{Department of Physics \& Astronomy, University of Wisconsin-Milwaukee, PO Box 413, WI 53201, Milwaukee, USA}
\email[A.~Majumder]{aishani@uwm.edu}

\author{Laura N. Driessen}
\affiliation{Sydney Institute for Astronomy, School of Physics, The University of Sydney, New South Wales 2006, Australia}

\author{Ashna Gulati}
\affiliation{Sydney Institute for Astronomy, School of Physics, The University of Sydney, New South Wales 2006, Australia}
\alsoaffiliation{ARC Centre of Excellence for Gravitational Wave Discovery (OzGrav), Australia}

\author{Tara Murphy}
\affiliation{Sydney Institute for Astronomy, School of Physics, The University of Sydney, New South Wales 2006, Australia}
\alsoaffiliation{ARC Centre of Excellence for Gravitational Wave Discovery (OzGrav), Australia}

\author{Dougal Dobie}
\affiliation{Sydney Institute for Astronomy, School of Physics, The University of Sydney, New South Wales 2006, Australia}
\alsoaffiliation{ARC Centre of Excellence for Gravitational Wave Discovery (OzGrav), Australia}

\received {13 03 2026}
\revised  {26 06 2026}
\accepted {12 08 2026}
\published {}

\doi{}
\begin{document}

\begin{abstract}

We present a search for radio emission from classical novae at $887.5\,\mathrm{MHz}$ using data from the Australian SKA Pathfinder Variable And Slow Transient (VAST) survey. We cross-matched 43 optically discovered classical novae that erupted between 2021 September and 2025 November within the $1200\,\mathrm{deg}^2$ Galactic survey footprint, and found three which show significant radio emission: V6598 Sgr, V1716 Sco, and V1723 Sco. To analyse their radio light curves, we use both thermal free-free and non-thermal synchrotron emission models. We fit the data using the Markov chain Monte Carlo (MCMC) method to constrain parameters, including the ejected mass and ejecta velocities for the thermal models, and mass-loss rate, explosion energy, wind velocity, and filling factor for the non-thermal model. All three novae show evidence of non-thermal synchrotron emission as the dominant emission mechanism at this frequency. We use a broken power law to describe the radial density structure of a non-uniform circumbinary material, which provides a better fit than a standard wind density profile. This strong early-time synchrotron emission is strong evidence of shock-driven particle acceleration, which may be related to detections of gamma-rays from all three novae as well. In contrast to earlier studies that used multi-frequency data to distinguish between emission models, our analysis is based on single-frequency radio light curves, which can still provide useful constraints on the dominant emission mechanism when interpreted with physically motivated models.

\end{abstract}

\section{Introduction}

\par
Recent radio observations have revealed a rich diversity in behaviour from the population of eruptive transients known as classical novae \citep[CNe;][]{2021ARA&A..59..391C,2021ApJS..257...49C}. A classical nova is a thermonuclear transient in a compact, accreting binary system known as a cataclysmic variable. In a semi-detached binary system of a white dwarf and a companion star, when the companion star overflows the Roche lobe, it starts to transfer its hydrogen-rich material through the inner Lagrangian point (L1). These materials accumulate on the surface of the white dwarf, and as the temperature and density increase, the innermost layer soon becomes degenerate. These conditions trigger an unstable thermonuclear runaway, igniting thermonuclear fusion \citep{1978ApJ...222..600S}. The outflow velocities from the white dwarf are typically in the range of hundreds to several thousand kilometres per second \citep{1978ARA&A..16..171G}. The total energy output ranges from $10^{38}$ to $10^{44}\,\mathrm{erg}$, enriching the surrounding interstellar medium with heavier materials \citep{2017MNRAS.470..401R}.

\par
Following the peak optical brightening at the mass ejection stage \citep{2011ApJS..194...28K}, the remnant envelope continues to support the nuclear burning, resulting in multi-component emission with complex radio light curves at different frequencies. Free-free thermal emission \citep{2012ApJ...761..173C} has traditionally been thought to dominate the radio emission from classical novae, but non-thermal synchrotron emission powered by the shock wave is now more commonly seen \citep{2012BaltA..21...62H}, especially from the gamma-ray producing novae studied by \citet{2014Sci...345..554A, 2021ARA&A..59..391C}. The evolution begins with the fireball expansion phase, during which the ejecta remain dense and optically thick due to absorption at all radio frequencies. During this phase, optically thick thermal emission from the photosphere dominates, resulting in a peak in the emission. The system becomes optically thin as density drops; that is, the photosphere recedes inward and traces the entire ejecta volume \citep{2012clno.book.....B}. The frequency at which the maximum amount of radiation escapes and the flux density drops further with expanding ejecta is called the turnover frequency. Following the peak, the flux density declines until the entire ejecta becomes fully transparent \citep{2023PASA...40...25G}.

\par
Most radio studies of CNe today have been follow-up observations \citep{2021ApJS..257...49C} of novae discovered at optical (or increasingly near-infrared; \citealt{2020MNRAS.492.4847L}) wavelengths. This can yield high-quality multi-frequency observations but is limited to sources that have attracted specific interest from radio observers and may also be biased toward novae that peak sooner (otherwise, follow-up efforts may cease). However, with the advent of high-cadence wide-field radio surveys \citep{2026PASA...43....6M}, serendipitous nova observations can cover a wider range of sources \citep[e.g.][]{2023PASA...40...25G}, albeit often lacking multi-frequency coverage. In this paper, we use the first $3.5\,\mathrm{years}$ of the Australian SKA Pathfinder \citep[ASKAP;][]{2021PASA...38....9H} Variables and Slow Transients \citep[VAST;][]{2013PASA...30....6M,2021PASA...38...54M} survey data covering the Galactic plane at $2\,\mathrm{week}$ cadence to identify and study a flux-limited sample of classical novae. The paper is organised as follows. Section \ref{sec:sample selection} outlines the source selection and summarises background information for the detected novae. Section \ref{sec:Analysis and Modelling} describes our approach to radio emission modelling, considering several emission models. In Section \ref{sec:results} we present the results of the light curve modelling. Finally, Section \ref{sec:discussion} summarises the findings with the conclusion presented in Section \ref{sec:conclusion}.

\subsection{The ASKAP Variables and Slow Transients Survey}

\par
ASKAP is a radio survey telescope operating with 36 antennas with a total instantaneous field of view $\approx31\,\mathrm{deg^2}$. The VAST survey conducted with ASKAP covers $1200\,\mathrm{deg}^2$ of the southern Galactic plane every two weeks with an integration time $\sim12$ to $15\,\mathrm{minutes}$ per pointing. The typical rms noise is $0.2\,\mathrm{mJy}$ with central frequency $887.5\,\mathrm{MHz}$ and bandwidth $288\,\mathrm{MHz}$. These high-cadence observations are optimised for monitoring transient and variable sources on timescales ranging from seconds to years. Data for individual sources are collected and calibrated as part of the VAST pipeline \citep{2022ASPC..532..333P}. Raw observational data for radio light curves are available in CASDA\footnote{See\url{https://research.csiro.au/casda/}.}

\section{Sample Selection}
\label{sec:sample selection}

\par
To search for classical novae, we began with the publicly available updated `Koji's list of recent Galactic Novae'\footnote{See \url{https://asd.gsfc.nasa.gov/Koji.Mukai/novae/novae.html}.}, which compiles spectroscopically confirmed novae across the Galactic plane since 2008. The novae are grouped by the year of their earliest detected optical eruption. As of the last update at the time of writing (6 January 2026), the catalogue contains approximately 188 confirmed novae, with the most recent detection on 29 November 2025. We cross-matched the novae with ASKAP VAST survey data processed through the VAST pipeline using a 5 arcsecond matching radius \citep{2023PASA...40...25G}, and searched for detectable radio emission for each candidate. 

\par
\begin{figure*}[t]
    \centering
    \includegraphics[width=0.49\textwidth]{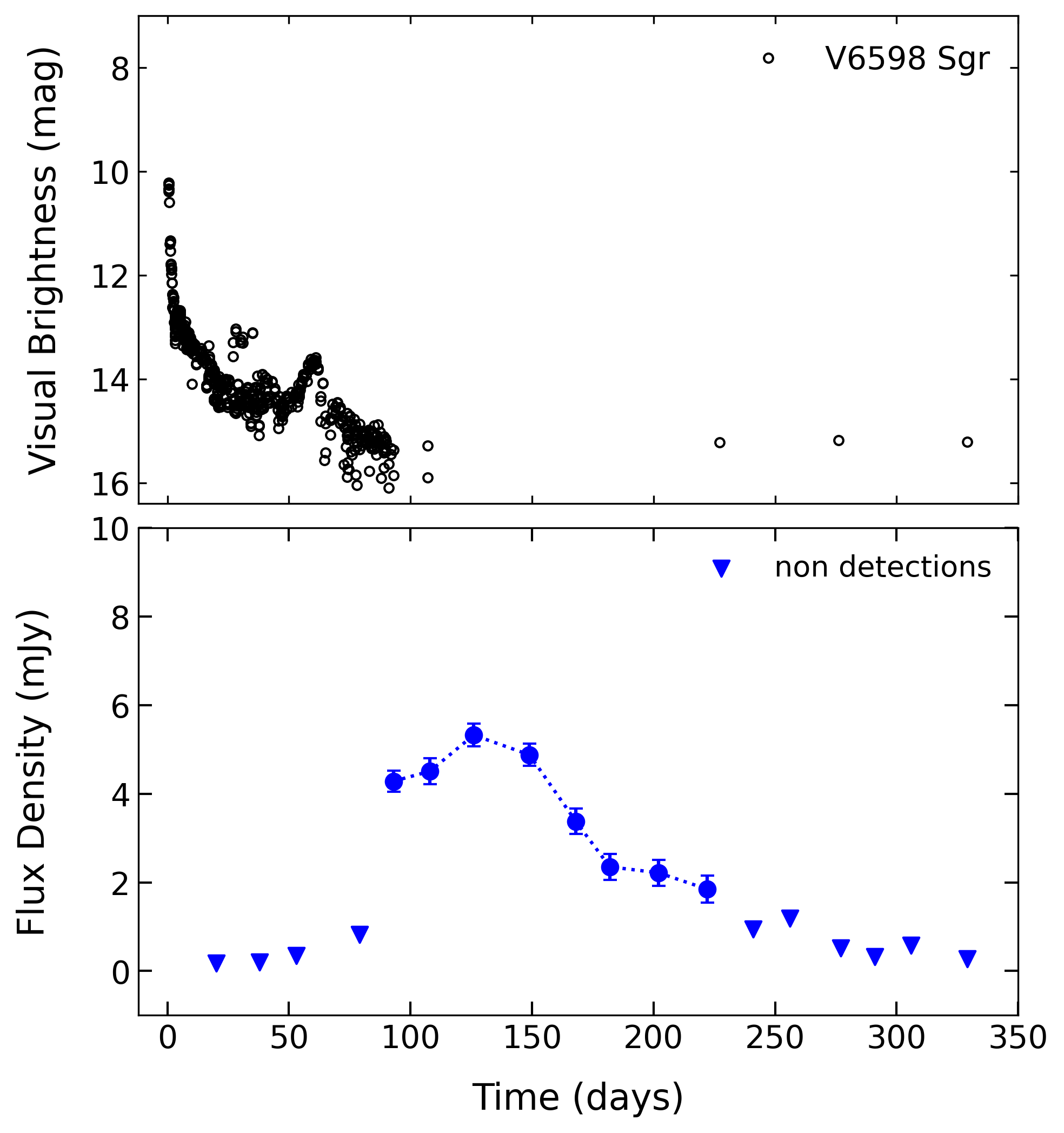}
    \hfill
    \includegraphics[width=0.49\textwidth]{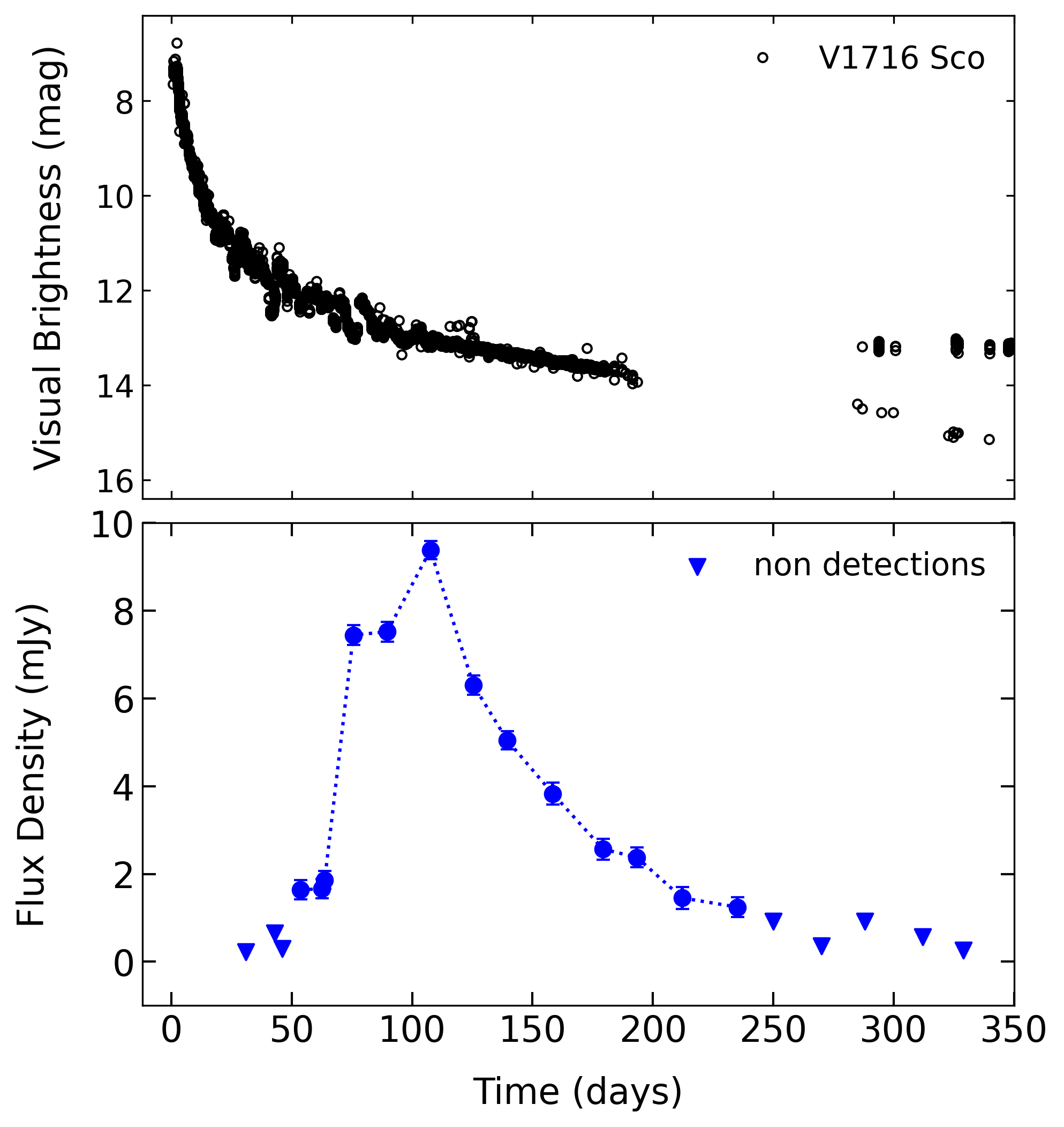}

    \vspace{0.4cm}

    \includegraphics[width=0.60\textwidth]{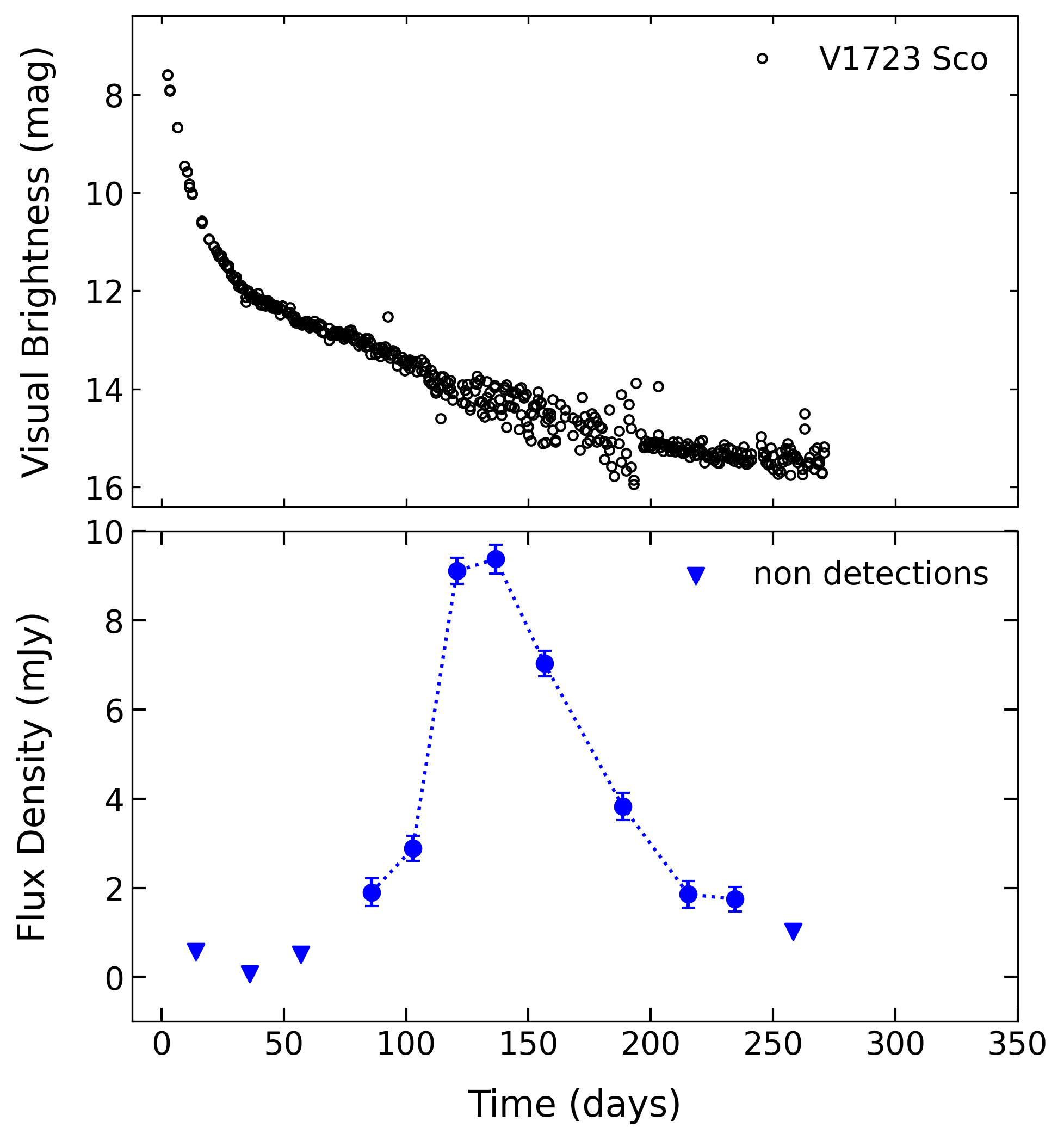}

    \caption{Optical (top) and radio (bottom) light curves for V6598 Sgr, V1716 Sco, and V1723 Sco. Optical data are from AAVSO; radio data at $887.5\,\mathrm{MHz}$ are from ASKAP VAST. Triangles indicate $5\sigma$ upper limits.}
    \label{fig:lightcurves}
\end{figure*}

The VAST survey commenced on 2022 January 1, and our nova sample includes 43 Galactic novae reported between 2021 September 10 ($\approx 100\,\mathrm{days}$ before January 2022) and 2025 November 29, of which 32 fell within the VAST Galactic survey footprint and were searched for radio emission. Among these, three of the classical novae showed well-sampled flux density evolution (discussed in detail below), while the remaining 29 were consistent with non-detections (or one or two points above the noise floor, insufficient for detailed analysis). The radio data on the detected sources are listed in Table~\ref{tab:data}. Optical light curves were produced by using the American Association of Variable Star Observers (AAVSO) Database in the Johnson V or Visual band, given by \citet{1993Ap&SS.210..137P}. Radio and optical light curves are shown together in Figure \ref{fig:lightcurves}, where we have taken the literature-reported discovery date as $t_0$. In the next sections, we summarise the previously reported multi-wavelength observations of individual novae, focusing on the most recent eruptions and data relevant to our modelling and interpretation.

\afterpage{%
\clearpage
\onecolumn
\color{black}
\footnotesize
\begin{longtable}{lcccccc}

\caption{ASKAP observations of classical novae from the VAST survey. Column 1 gives the General Catalogue of Variable Stars (GCVS) ID \citep{1981PASP...93..165D} of the novae. Columns 2 and 3 are right ascension and declination. Column 4 is the discovery date, which we considered as $t_0$ in optical detection. Column 5 contains the date of radio observations conducted with ASKAP, and Column 6 contains the number of days from the detection date. Column 7 gives the integrated flux density in $\mathrm{mJy}$ for $\geq5\sigma$ detections or the $5\sigma$ upper limit.}
\label{tab:data}\\

\toprule
\shortstack{GCVS\\ID} &
\shortstack{RA\\(J2000)} &
\shortstack{Dec\\(J2000)} &
\shortstack{$t_0$\\(UTC)} &
\shortstack{Observation date\\(UTC)} &
\shortstack{Time\\($\mathrm{days}$)} &
\shortstack{$S_{\mathrm{int}}$\\($\mathrm{mJy}$)} \\
\midrule
\endfirsthead

\toprule
\shortstack{GCVS\\ID} &
\shortstack{RA\\(J2000)} &
\shortstack{Dec\\(J2000)} &
\shortstack{$t_0$\\(UTC)} &
\shortstack{Observation date\\(UTC)} &
\shortstack{Time\\($\mathrm{days}$)} &
\shortstack{$S_{\mathrm{int}}$\\($\mathrm{mJy}$)} \\
\midrule
\endhead

\midrule
\multicolumn{7}{r}{\textit{Continued on next page}}\\
\endfoot

\bottomrule
\endlastfoot

V6598 Sgr & 17:52:49.30 & $-20$:24:15.5 & 2023-07-15 & 2023-08-04 & 20 & $<1.10$ \\
&&&& 2023-08-22 & 38 & $<1.55$ \\
&&&& 2023-09-06 & 53 & $<1.20$ \\
&&&& 2023-10-02 & 79 & $<1.10$ \\
&&&& 2023-10-16 & 93 & $4.3\pm0.24$ \\
&&&& 2023-10-31 & 108 & $4.5\pm0.29$ \\
&&&& 2023-11-18 & 126 & $5.3\pm0.26$ \\
&&&& 2023-12-11 & 149 & $4.8\pm0.25$ \\
&&&& 2023-12-30 & 168 & $3.3\pm0.29$ \\
&&&& 2024-01-13 & 182 & $2.3\pm0.29$ \\
&&&& 2024-02-02 & 202 & $2.2\pm0.29$ \\
&&&& 2024-02-22 & 222 & $1.8\pm0.31$ \\
&&&& 2024-03-11 & 241 & $<1.50$ \\
&&&& 2024-03-26 & 256 & $<1.20$ \\
&&&& 2024-04-16 & 277 & $<1.20$ \\
&&&& 2024-04-30 & 291 & $<1.35$ \\
&&&& 2024-05-15 & 306 & $<1.20$ \\
&&&& 2024-06-07 & 329 & $<1.40$ \\
\midrule

V1716 Sco & 17:22:45.05 & $-41$:37:16.3 & 2023-04-20 & 2023-05-21 & 31 & $<1.0$ \\
&&&& 2023-06-02 & 43 & $<1.15$ \\
&&&& 2023-06-05 & 46 & $<1.15$ \\
&&&& 2023-06-12 & 53 & $1.6\pm0.2$ \\
&&&& 2023-06-21 & 62 & $1.7\pm0.2$ \\
&&&& 2023-06-22 & 63 & $1.8\pm0.2$ \\
&&&& 2023-07-04 & 75 & $7.4\pm0.2$ \\
&&&& 2023-07-18 & 89 & $7.5\pm0.2$ \\
&&&& 2023-08-05 & 107 & $9.3\pm0.2$ \\
&&&& 2023-08-23 & 125 & $6.3\pm0.2$ \\
&&&& 2023-09-06 & 139 & $5.0\pm0.2$ \\
&&&& 2023-09-25 & 158 & $3.8\pm0.2$ \\
&&&& 2023-10-16 & 179 & $2.5\pm0.2$ \\
&&&& 2023-10-30 & 193 & $2.3\pm0.2$ \\
&&&& 2023-11-18 & 212 & $1.4\pm0.3$ \\
&&&& 2023-12-11 & 235 & $1.2\pm0.2$ \\
&&&& 2023-12-26 & 250 & $<1.40$ \\
&&&& 2024-01-15 & 270 & $<1.20$ \\
&&&& 2024-02-02 & 288 & $<1.30$ \\
&&&& 2024-02-26 & 312 & $<1.20$ \\
&&&& 2024-03-14 & 329 & $<1.20$ \\
\midrule

V1723 Sco & 17:26:18.09 & $-38$:09:36.3 & 2024-02-08 & 2024-02-21 & 14 & $<2.0$ \\
&&&& 2024-03-14 & 36 & $<1.35$ \\
&&&& 2024-04-04 & 57 & $<1.35$ \\
&&&& 2024-05-03 & 86 & $1.9\pm0.3$ \\
&&&& 2024-05-20 & 103 & $2.8\pm0.3$ \\
&&&& 2024-06-07 & 121 & $9.1\pm0.3$ \\
&&&& 2024-06-23 & 137 & $9.3\pm0.3$ \\
&&&& 2024-07-13 & 157 & $7.0\pm0.3$ \\
&&&& 2024-08-14 & 189 & $3.8\pm0.3$ \\
&&&& 2024-09-10 & 215 & $1.8\pm0.30$ \\
&&&& 2024-09-29 & 234 & $1.7\pm0.28$ \\
&&&& 2024-10-23 & 258 & $<1.20$ \\

\end{longtable}

\twocolumn
\normalsize
\clearpage
}%

\subsection{V6598 Sgr}

\par
V6598 Sgr was reported in the visible band by Andrew Pearce and Y.~Nakamura \citep{2023CBET.5278....1P} on 2023 July 15, following a rapid rise of $>3\,\mathrm{mag}$ within $<24\,\mathrm{hours}$. Around the same time, it was detected by the \textit{Fermi} Gamma-ray Observatory. The initial brightness was already near the maximum of $10\,\mathrm{mag}$ \citep{2023ATel16141....1M} and dropped to $12.5\,\mathrm{mag}$ in $\sim 2\,\mathrm{days}$ \citep{2023CBET.5245....1P}. Spectral analysis detected broad H$\alpha$ emission with a maximum projected velocity $4000\,\mathrm{km\,s^{-1}}$ and from P-Cyg absorption line $3500\,\mathrm{km\,s^{-1}}$ decreasing from $3700$ to $3300\,\mathrm{km\,s^{-1}}$, leading to a classification as a fast nova with powerful ejecta with following a deceleration \citep{2023ATel16141....1M}.

\par
X-ray observations were obtained by \citet{2021ApJ...914...85H} using the \textit{Neil Gehrels Swift Observatory} \citep{2004ApJ...611.1005G}. Based on these observations, the system was classified as an intermediate polar (IP). The early evolution of V6598 Sgr shows similarities to two other He/N novae, V2672 Oph (Nova Oph 2009) and U Sco \citep{2023ATel16141....1M}. X-ray emission from U Sco resumed after $\sim 9\,\mathrm{days}$ of optical outburst, whereas V6598 Sgr re-emerged in X-rays after $\sim 17\,\mathrm{days}$, suggesting that this is a magnetic binary system in which the accretion process resumed after its optical discovery \citep{2023ATel16172....1N}.

\par
Radio emission from V6598 Sgr was observed repeatedly from 2022 November 14 to 2024 October 10 at $887.5\,\mathrm{MHz}$ in the VAST survey and reported by \citet{2023ATel16383....1D}. The source was detected above the $\geq5\sigma$ upper limit in 8 epochs between 2023 October 16 and 2024 February 22. It was subsequently detected at $944\,\mathrm{MHz}$ on 2024 January 15 and 26 by \citet{2025ATel16969....1R} in the Rapid ASKAP Continuum Survey \citep[RACS;][]{2020PASA...37...48M}. A steady increase in flux density from $4.3\pm0.2$ to $5.3\pm0.2\,\mathrm{mJy}$ between day 93 and 126 was observed \citep{2023ATel16383....1D}. 

The geometric (purely parallax-based) and photogeometric (including priors to account for Lutz-Kelker bias; \citep{1973PASP...85..573L}) distances are $5.8^{+1.96}_{-1.97}\,\rm{kpc}$ and $4.3^{+1.92}_{-1.18}\,\rm{kpc}$ \citep{2021yCat.1352....0B} respectively, neither of which is statistically significant, and where the priors used for those photogeometric distances are not necessarily appropriate for CNe. In a recent study by \citet{2025ApJ...993..232S}, the distance of V6598 Sgr is estimated to be $7.661\,\mathrm{kpc}$ based on the Bayesian distance analysis \citep{2022MNRAS.517.6150S}, which uses Gaia parallaxes together with priors appropriate for Galactic nova populations and previous distance estimates with uncertainties. In this work, we use a distance of $7.6\,\rm{kpc}$ for our further model fitting and parameter estimations.

\subsection{V1716 Sco}

\par
V1716 Sco was discovered by \citet{2023CBET.5245....1P} on 2023 April 18 and 19 with a rapid rise from $12.5$ to $7.3\,\mathrm{mag}$ within two days \citep{2023ATel16018....1S}. Later, it decreased by $2$ to $3\,\mathrm{mag}$ in the next $15$ to $20\,\mathrm{days}$, making it a very fast nova according to the speed classes of classical novae \citep{2010AJ....140...34S}. Between $150$ and $200\,\mathrm{days}$, the visual brightness settled at $14$ to $15\,\mathrm{mag}$ and returned to its quiescent state. V1716 Sco hosts a massive white dwarf of $1.21\,M_{\odot}$ with a main-sequence star as donor \citep{2025ApJ...993..232S}, exhibiting typical behaviour of fast novae \citep{2002ASPC..261..595K}. This is confirmed as a Fe\,II classical nova by spectroscopic analysis \citep{2024ApJ...968...31W}. Spectra obtained by the Southern Spectroscopic Observatory Project (2SPOT) revealed complex and strong Balmer P-Cyg absorption lines with velocities of $1800\,\mathrm{km\,s^{-1}}$ using the Deep Sky Chile Telescope \citep{2023CBET.5245....1P}.

\par
After the optical outburst, X-rays and gamma-rays were detected within a day by NICER, NuSTAR, and \textit{Fermi}-LAT \citep{2023ATel16018....1S}. The X-ray was observed from an optically thin thermal plasma heated by shocks, with a peak at $\sim 20\,\mathrm{days}$ after the eruption, entering the super soft (SSS) phase about $55\,\mathrm{days}$ later. Consistent with behaviour observed in many classical novae, detection of early X-rays and $\gamma$-rays implies a slow, low-density outflow dominated by a faster and denser wind, producing shocks \citep{2024arXiv240619233W}. During the SSS phase \citet{2025ApJ...995...30W} reported a $78\,\mathrm{s}$ quasi-periodic oscillation (QPO) in X-rays with NICER, indicating variability in the emitting region with ongoing nuclear burning. From recently released Transiting Exoplanet Survey Satellite (TESS) archival data, the orbital period is measured to be $1.36\,\mathrm{days}$ following an exceptionally low ratio of $P_{\mathrm{spin}} / P_{\mathrm{orb}}=6.6 \times 10^{-4}$ \citep{2025arXiv251106399L}.

\par
V1716 Sco was observed in multiple epochs between 2022 November 19 and 2024 October 3 as part of the VAST survey. Significant radio emission was detected in 13 epochs between 2023 June 12 and 2023 December 11 \citep{2023ATel16155....1A}. The radio flux density rose rapidly, reaching a peak of $9.3\pm0.2\,\mathrm{mJy}$ on day 107, before declining. The optical peak precedes the radio rise by $60$ to $70\,\mathrm{days}$. The geometric  and photogeometric distances \citep{2021AJ....161..147B} are  $3.19^{+2.13}_{-1.62}\,\mathrm{kpc}$ and $4.96^{+1.75}_{-1.06}\,\mathrm{kpc}$ respectively \citep{2024arXiv240619233W}. Based on the Bayesian distance analysis of \citet{ 2025ApJ...993..232S}, which includes available \textit{Gaia} information along with other data specific to novae, the reported distance for this source is $2.914\,\rm{kpc}$, which is close to the geometric estimate of \citet{2021AJ....161..147B}. Given the large uncertainty, we adopted a distance of $3\,\mathrm{kpc}$. However, we also evaluated how the inferred parameters scale with distance to highlight the uncertainty.

\subsection{V1723 Sco}

\par
V1723 Sco was discovered on 2024 February 8 \citep{2024ATel16444....1S} with rapidly rising brightness, reaching a maximum value of $6.77\,\mathrm{mag}$ in \textit{V}-band \citep{2024ATel16454....1H} within a few days of the optical eruption. At the time of discovery, the visual brightness was still increasing, and by day $\sim4$ it had declined to $8.1\,\mathrm{mag}$, remaining at that level for the next few days. From the two optical spectra reported by 2SPOT on 2024 February 11 and 12, the nova shell appear to be still expanding at $3200\,\mathrm{km\,s^{-1}}$ (H$\alpha$ line) and $1800\,\mathrm{km\,s^{-1}}$ (H$\beta$ line) respectively \citep {2024ATel16454....1H}. 

\par
Gamma-ray emission above $100\,\mathrm{MeV}$ was detected starting on 2024 February 9 \citep{2024ATel16439....1C}. The follow-up observations revealed an average photon flux of  $(1.9\pm0.5) \times 10^{-6}\,\mathrm{photons\,cm^{-2}\,s^{-1}}$ on 2024 February 11 to 12, $\sim 2$ to $3\,\mathrm{days}$ after the outburst. The typical range of photon flux for classical novae reported by \textit{Fermi}-LAT is $10^{-7}\,\mathrm{photons\,cm^{-2}\,s^{-1}}$ \citep{2014Sci...345..554A}. Novae such as V1324 Sco \citep{2018ApJ...852..108F}, V5668 Sgr \citep{2016ApJ...826..142C} are example of the brighter end of this population. Only a few classical novae including V407 Cyg \citep{2010Sci...329..817A}, V959 Mon \citep{2018ApJ...858..108H} have reached such brightness ($10^{-6}\,\mathrm{photons\,cm^{-2}\,s^{-1}}$) similar to V1723 Sco. Therefore, this is one of the strongest and most luminous gamma-ray-detected classical novae since 2010 \citep{2024ATel16441....1C}. This source was observed from February 10.299-11.767 UT in hard X-rays by NuSTAR and in soft X-rays by Swift/XRT, up to February 13, with both showing non-detections \citep{2024ATel16444....1S}. Later, on 2024 March 13 ($33\,\mathrm{days}$ from the optical outburst), it reached its peak in the soft band (0.3--2\,$\mathrm{keV}$), and became one of the highest level of shock-powered classical novae to reach such X-ray brightness \citep{2026A&A...705A..19F}.   

\par
Radio observations of V1723 Sco started on 2024 February 17 using the Karl G.\ Jansky Very Large Array (VLA) Telescope with a magnitude of $7.8\,\mathrm{mJy}$ \citep{2024ATel16492....1M}. The calculated spectral index from multi-frequency data is $\alpha=0.5$, suggesting an optically thick phase in its radio flux evolution. For V1723 Sco we used a distance of $\sim 8\,\rm{kpc}$ from \citet{2025ApJ...993..232S}, adopting the  Bayesian analysis from \citet{2022MNRAS.517.6150S}. From 2024 February 21 to April 4, $\sim 14$ to $57\,\mathrm{days}$ after the eruption, this classical nova candidate remained undetected with a $3\sigma$ upper limit in all available epochs of the VAST survey. The first detection by ASKAP was made $86\,\mathrm{days}$ post-eruption, on 2024 May 3, at $1.9\pm0.3\,\mathrm{mJy}$. The source peaked on day 137 and was last detected at $1.7\pm0.28\,\mathrm{mJy}$, after which it became undetected.

\section{Analysis and Modelling}
\label{sec:Analysis and Modelling}

To fit the data, we considered both thermal (bremsstrahlung) \citep{1979AJ.....84.1619H,1989agna.book.....O} and non-thermal (magnetobremsstrahlung) \citep{1998ApJ...499..810C,2023MNRAS.523.1661N} emission models. To estimate the uncertainties in flux density, we used the quadrature sum of the local RMS, peak-flux error, and $10\%$ flux-scaling error, following \citet{McConnell2015_ASKAPapertureEfficiency}. For quantitative assessment of the quality of each fit, we used the Akaike Information Criterion (AIC; \citealt{2012MNRAS.419.3292T}) and the Bayesian Information Criterion (BIC; \citealt{2007MNRAS.377L..74L}). Here, we briefly discuss each of our emission models. We present one corner plot for each source with the lowest reduced $\chi^2$ value, which also corresponds to the lowest AIC and BIC values.

\subsection{Free-Free thermal emission}
\par
After the nova eruption, radio emission may arise from the hot, ionised plasma undergoing Hubble-type expansion away from the white dwarf. The intensity of this emission depends on the electron temperature and plasma density. The radio light curve represents the evolution of the optical depth $\tau_\nu$ along with the expansion of the ejecta. It initially rises during an optically thick phase ($\tau_\nu \gg 1$), reaches a peak when the optical depth of the ejecta approaches unity, and then declines further as the ejecta expands ($\tau_\nu \ll 1$) and becomes optically thin \citep{1989agna.book.....O, 2008clno.book..141S,2003AJ....125..465H}. We adopted the standard free-free thermal model with a linear velocity gradient for the nova shell, as given by \citet{1979AJ.....84.1619H}. The free parameters are the ejecta mass $M_{\mathrm{ej}}$, inner and outer velocities $v_1$ and $v_2$, inner and outer radii $r_1$ and $r_2$, and electron temperature $T_{\mathrm{e}}$. In the optically thin limit, the free-free flux density can be defined as,

\begin{equation}
    S_{\nu_{\mathrm{thin}}}=C \left[\frac{M_{\mathrm{ej}}^2}{d^2\,r_2\,r_1\,(r_2-r_1)}\right]
\end{equation}

where the coefficient $C$ is;

\begin{equation}
    C = \frac{J_0}{4\pi\,m_p^2}\, T_e^{-0.5}\, g_{ff}\, \exp \left[-\frac{h\nu}{k_B T_e}\right]
\end{equation}

\par
In the expression above, $J_0$ is the free-free normalisation constant and $g_{\rm ff}$ is the free-free Gaunt factor, both defined from \citet{1989agna.book.....O}, and $m_p$ is the proton mass. Considering $T_e=10^4\,\rm{K}$ and $\nu=887.5\,\rm{MHz}$, the value of this constant in cgs unit is $C=9.32\times10^6\,\mathrm{erg\,\,s^{-1}\,Hz^{-1}cm^{3}\,g^{-2}}$. In the light curve fitting, we used the full expression, including free-free self-absorption, allowing the model to transition between optically thick and optically thin regimes. For Hubble-like expansion $r_1$ and $r_2$ are given by:

\begin{equation}
     r_1 = r_{10} + v_1(t-t_0) 
\end{equation}
and
\begin{equation}
         r_2 = r_{20} + v_2(t-t_0) 
\end{equation}

\par
At a time $t$ compared to the eruption time $t_0$. At $t_0=0$, the initial radii of the ejected shell are $r_{10}$ and $r_{20}$, although we ignore these in subsequent analysis (since the radio emission is at a late time with $r\gg r_0$). We fixed the electron temperature at $10^4\,$K since the dependence of the emission on $T_{\mathrm{e}}$ is very weak when just dealing with a single radio frequency.

\par
The model then becomes a three-parameter model with parameters $M_{\mathrm{ej}}$, $v_2$, and $v_1/v_2$. We fit the data with the MCMC method by using the \texttt{emcee} package \citep{2013PASP..125..306F}. Based on the parameter values commonly reported in previous studies of classical novae, as referenced above, we set the priors as;  $10^{-7}\,\mathrm{M_{\odot}} < M_{\mathrm{ej}} < 10^{-4}\,\mathrm{M_{\odot}}$, $100\,\mathrm{km\,s^{-1}} < v_2 < 10000\,\mathrm{km\,s^{-1}}$ and $0.4 < v_1/v_2 < 0.8$. Our fits used 60 walkers over 1000 steps. We produced corner plots to visualise the posterior distributions and best-fit values with uncertainties, and used them to generate sample models for visual comparison with the observational data.

\subsection{Non-thermal synchrotron emission}
\par
When the fast, rapidly expanding ejecta interact with slower, less dense circumbinary material, it generates a shock wave \citep{1992ApJ...400..222B} at the interface of the two media. The interaction between the faster ejecta and the earlier and slower outflow from the white dwarf surface concentrated at the equatorial plane of the binary produces internal shocks typically observed in classical novae \citep{2014MNRAS.442..713M}. The collision between fast ejecta and dense stellar wind from the companion produces external shocks associated with symbiotic and recurrent novae \citep{2016MNRAS.463..394V}. Both shock waves carry the energy of the outburst and transfer a fraction ($\epsilon_{\mathrm{e}}$) of this energy to accelerate electrons to relativistic speeds and a fraction ($\epsilon_{\mathrm{B}}$) to amplify the magnetic field. Non-thermal synchrotron emission arises when these relativistic electrons spiral along magnetic field lines. This energy transfer process follows a power-law energy distribution: $N(E)\,dE \propto E^{-p}\,dE$. Therefore, the light curve evolves based on the circumbinary density, the magnetic field strength, the change in optical depth because of self-absorption (SSA; \citealt{1994ApJ...420..268C}) or free-free absorption (FFA), the geometry of the ejecta, and the area of the emitting region \citep{1998ApJ...499..810C}.

\par
From the point of discontinuity, the forward shock propagates outward, and the reverse shock propagates inward through the faster ejecta. The synchrotron-emitting region lies within this shock and the contact discontinuity, with a volume-filling factor $f\leq 1$. Emission from the reverse shock is usually absorbed by the cold central shell \citep{2016MNRAS.463..394V}, which dominates at frequencies below some turnover. Above the turnover frequency, the light curve transitions from an optically thick to an optically thin phase $\tau_\nu\ll1$. To fit this type of emission, we adopted the standard synchrotron model from \citet{2023MNRAS.523.1661N} and \citet{1998ApJ...499..810C}, where the surrounding material is approximated by an effective $\rho(R) \propto R^{-2}$ density profile, with $R$ being the expanding shock front radius. In applying this model to classical novae, we interpret this density profile as an effective description of the slow ejecta or circumbinary material encountered by the shock. To further incorporate this emission mechanism, we considered both synchrotron self-absorption (SSA) and an external free-free absorption (FFA) factor. While SSA regulates within the shocked region, FFA from ionised material along the line of sight further suppresses the emission and delays the radio peak. Here, we applied an intrinsic SSA with an external FFA screen to model our observational data.

\par
However, as we shall see, this single power-law model cannot explain the steep post-peak decline observed in these very fast, early synchrotron flares. Different classical novae exhibit distinct rise and decay slopes in their radio evolution, indicating that a single smooth density profile may not be sufficient to describe the effective density structure encountered by the shock. While some novae have been modelled with aspherical ejecta, for example, V959 Mon \citep{2017MNRAS.469.3976H}, RS Oph \citep{2022A&A...666L...6M}, V445 Pup \citep{2021MNRAS.501.1394N}, V407 Cyg \citep{2015ApJ...806...27P}, or with variations in filling factor, we needed a simpler approach given our limited single-frequency data. Therefore, to capture this complex, asymmetric temporal behaviour, we modified the density profile ($\rho$) of this shock-interacting material by introducing a broken power law. Hence, $\rho$ falls at a different rate in two separate regions while maintaining a smooth, continuous transition, i.e., at $R\ll R_0$ and $R\gg R_0$, where $R_0$ is the characteristic scale radius or the break point of the two different regions. A simple broken power law for this surrounding density structure can be written as

\begin{equation}
\rho \propto \left( \frac{R}{R_0} \right)^{-k}\left\{\frac{1}{2}\left[1 + \left( \frac{R}{R_0} \right)^{1/n}\right]\right\}^{(k - m)n}
\end{equation}

\par
We used three different power $k$, $m$ and $n$, where $k$ controls the inner slope, i.e, how steeply the density drops at $R\ll R_0$ region, $n$ controls the transition smoothness or how abruptly the density profile turns over near the break radius $R_0$ and lastly $m$ controls the steepness of post peak decay at large $R$ ($R\gg R_0$). Three sources showed different rates of steepness in their light curves. However, we did not fit for these indices, we fixed them at certain values depending on the steepness of temporal evolution as $R(t)\propto t$ then fit for five parameters such as filling factor $f$, wind velocity $V_{\mathrm{wind}}$, mass loss rate $\dot M$, explosion energy $E$, ejected mass $M_{\mathrm{ej}}$, via MCMC sampling. In this work, we considered the model parameter $V_{\rm{wind}}$ as the velocity of the slower circumbinary material that the shock encounters. We chose the fraction of energy transferred to electrons $\epsilon_{\mathrm{e}}=0.01$ and the fraction of energy transferred to the magnetic field $\epsilon_{\mathrm{B}}=0.01$ and $p=2.5$ for our calculations \citep{pacholczyk1970}. The best-fit model was sampled with 100 walkers over 1000 steps. Similarly based on the parameter values commonly reported in previous synchrotron emission studies of classical novae, as referenced above, the adopted priors for MCMC sampling are; $0.01 < f < 1$, $50\,\mathrm{km\,s^{-1}}<V_{\mathrm{wind}}<400\,\mathrm{km\,s^{-1}}$, $10^{-7}\,\mathrm{M_{\odot}}\,\mathrm{yr^{-1}} < \dot{M} < 10^{-4}\,\mathrm{M_{\odot}}\,\mathrm{yr^{-1}}$, $10^{42}\,\mathrm{erg} < E < 10^{45}\,\mathrm{erg}$ and $10^{-7}\,\mathrm{M_{\odot}} < M_{\mathrm{ej}} < 10^{-4}\,\mathrm{M_{\odot}}$. After the MCMC run, we took 100 samples with uncertainties and plotted them to visually compare with the data.

\subsection{Distance Scaling}

\par
For each source, we computed the best-fit model for the nominal distances in Section \ref{sec:sample selection}. We did not allow the distance to be a free parameter in the fit, since the fitted parameters are highly degenerate with distance. However, since many of the distances we adopt are themselves highly uncertain, we provide the scaling factors for each fitted parameter with respect to distance. In general, the flux density will scale roughly as a power-law function of each parameter (at least when close to the best-fit values), and to compensate for the $d^{-2}$ scaling of the flux density with distance, we derived the scaling indices that are listed in Tables \ref{tab:Distance Scaling_thermal} and \ref{tab: Distance Scaling_nonthermal}. 

\par
We started by calculating the power-law indices analytically for each parameter as a function of distance. However, for parameters like $E$, $\dot{M}$, $V_{\mathrm{wind}}$, and $M_{\mathrm{ej}}$ in the synchrotron model, the model is not a pure power law. The distance-scaling exponent itself varies with distance. Therefore, we estimated the local exponent at the flux density peak ($t_{\mathrm{peak}}\approx100\,\mathrm{days}$) by using the best-fit value and our reference distance.

\begin{table}[t]
  \centering
  \caption{Approximate power-law scaling indices for each parameter with distance $d$ for the free-free emission model. We kept the electron temperature fixed at $T_{\mathrm{e}}=10^4\,\mathrm{K}$.}
  \label{tab:Distance Scaling_thermal}
  \setlength{\tabcolsep}{2pt}
  \renewcommand{\arraystretch}{1.4}

  \begin{tabular*}{\textwidth}{@{\extracolsep{\fill}} l c c }
    \hline
    \textbf{Parameter (unit)} & \textbf{Scaling measurement} \\
    \hline

    Ejected mass $M_{\mathrm{ej}}\,(\mathrm{M}_{\odot})$ & $M_{\mathrm{ej}}\propto d$  \\

    Outer velocity $v_2\,$($\mathrm{km\ s^{-1}}$) & $v_2\propto d^{-2/3}$ \\

    \hline
  \end{tabular*}
\end{table}

\begin{table}[t]
  \centering
  \caption{Approximate power-law scaling indices for each parameter with distance $d$ for the synchrotron emission model, with electron power-law index fixed to $p=2.5$. }
  \label{tab: Distance Scaling_nonthermal}
  \begin{tabular*}{\textwidth}{@{\extracolsep{\fill}} l c }
    \hline
    \textbf{Parameter (unit)} & \textbf{Scaling measurement}\\
    \hline
    Filling factor $f$  & $f\propto d^{2}$  \\
    
    Wind velocity $V_{\mathrm{wind}}\,(\mathrm{km\,s^{-1}})$ & $V_{\mathrm{wind}}\,\propto\,d^{-1.28}$  \\
    
    Mass-loss rate $\dot{M}\,(\mathrm{M}_\odot\,\mathrm{yr^{-1}})$ & $\dot{M}\propto d^{1.28}$ \\
    
    Kinetic energy $E\,(\mathrm{erg})$ &  $E\propto d^{1.49}$   \\
    
    Ejecta mass $M_{\mathrm{ej}}\,(\mathrm{M}_{\odot})$  & $M_{\mathrm{ej}}\,\propto\,d^{-1.92}$    \\
    \hline
  \end{tabular*}
\end{table}

\section{Results}
\label{sec:results}

\subsection{V6598 Sgr}

\par

\begin{figure}[htbp]
  \centering
  \includegraphics[width=1\textwidth]{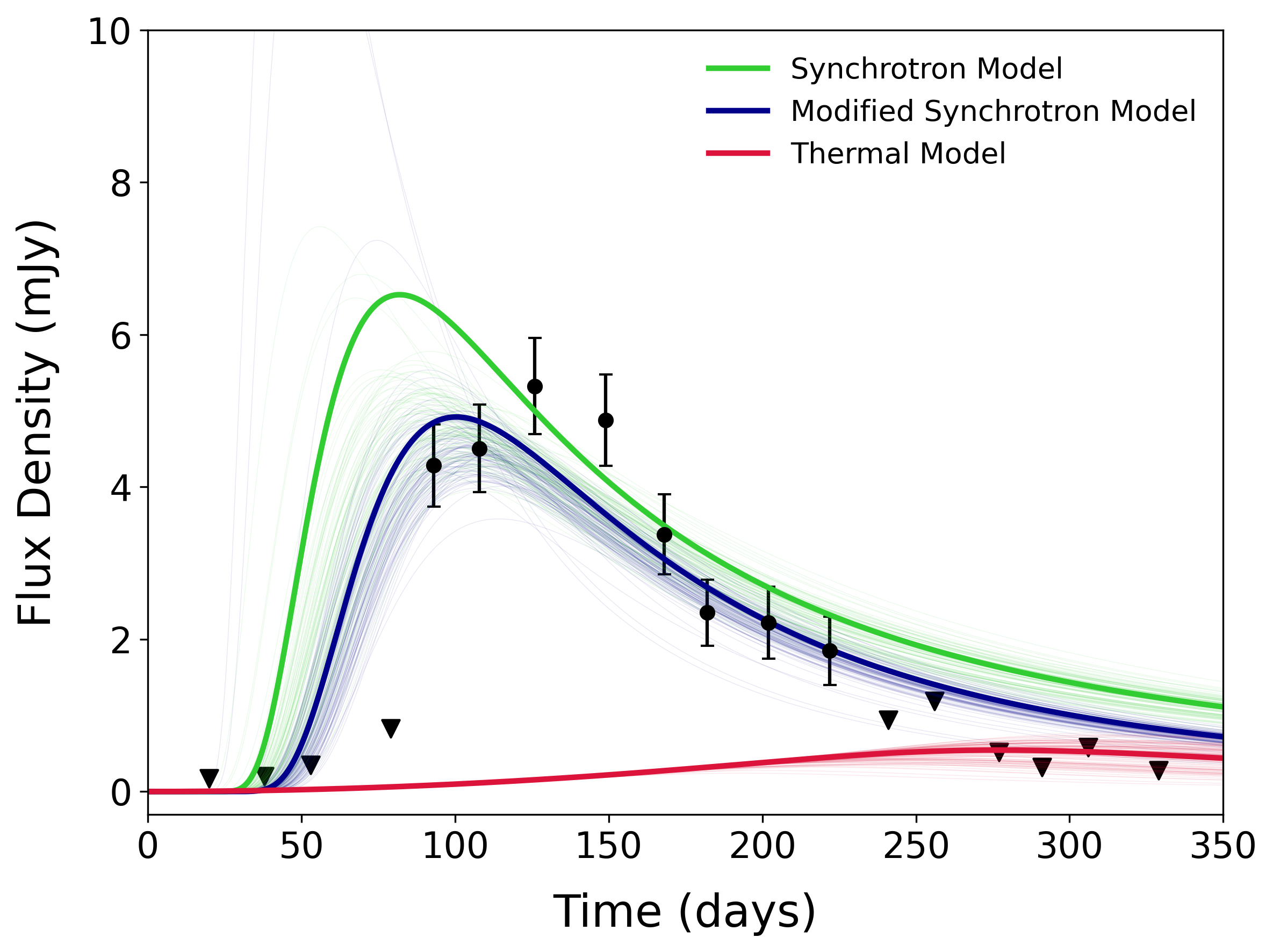} 
  \caption{Model fit of thermal, non-thermal, and modified non-thermal emission to V6598 Sgr at $887.5\,\mathrm{MHz}$, with the red, green, and blue light curves representing the respective model fits. Spaghetti plots based on 100 posterior samples from MCMC fitting are shown to illustrate the spread and uncertainty of the model, while the bold curve denotes the best-fit model.}
  \label{fig:V6598_ModelFit}
\end{figure}

We statistically compared the fit of all three models shown in Figure \ref{fig:V6598_ModelFit} and concluded that the modified synchrotron model provides a comparatively better fit to the observed data, with a reduced $\chi^2=2.9$ and 3 degrees of freedom. The best-fit values from MCMC analysis are listed in Table \ref{tab:Mod_Synchrotron_Model}. The thermal model rises too slowly and remains faint during the observed radio maximum. It suggests that a single spherically symmetric, uniformly filled thermal shell is unlikely to capture the dominant emission mechanism in this source. Instead, the observed light curve is more consistent with a shock-powered synchrotron emission component whose visibility evolves as the ejecta and the surrounding material become optically thin at $887.5\,\rm{MHz}$.

\par
This result is consistent with the multi-frequency evidence for shocks in V6598 Sgr, including the short-lived \textit{Fermi}-LAT detection, broad optical emission lines, and rapid early evolution. The radio emission peak indicates that the low-frequency synchrotron emission becomes visible on a $\sim4$ month timescale. From the light curve, we can observe that the transition from optically thick to thin is more gradual and smoother here than in the other two sources. In our synchrotron model, this behaviour is captured by the adopted density transition parameter $n = 0.7$. This may indicate that the shock is propagating through a comparatively dense or smoothly varying environment, consistent with the gamma-ray interpretation that the accelerated particles may be located in a dense region \citep{2026A&A...705A..19F}.

\par
\begin{figure*}[htbp]
  \centering
  \includegraphics[width=1\textwidth]{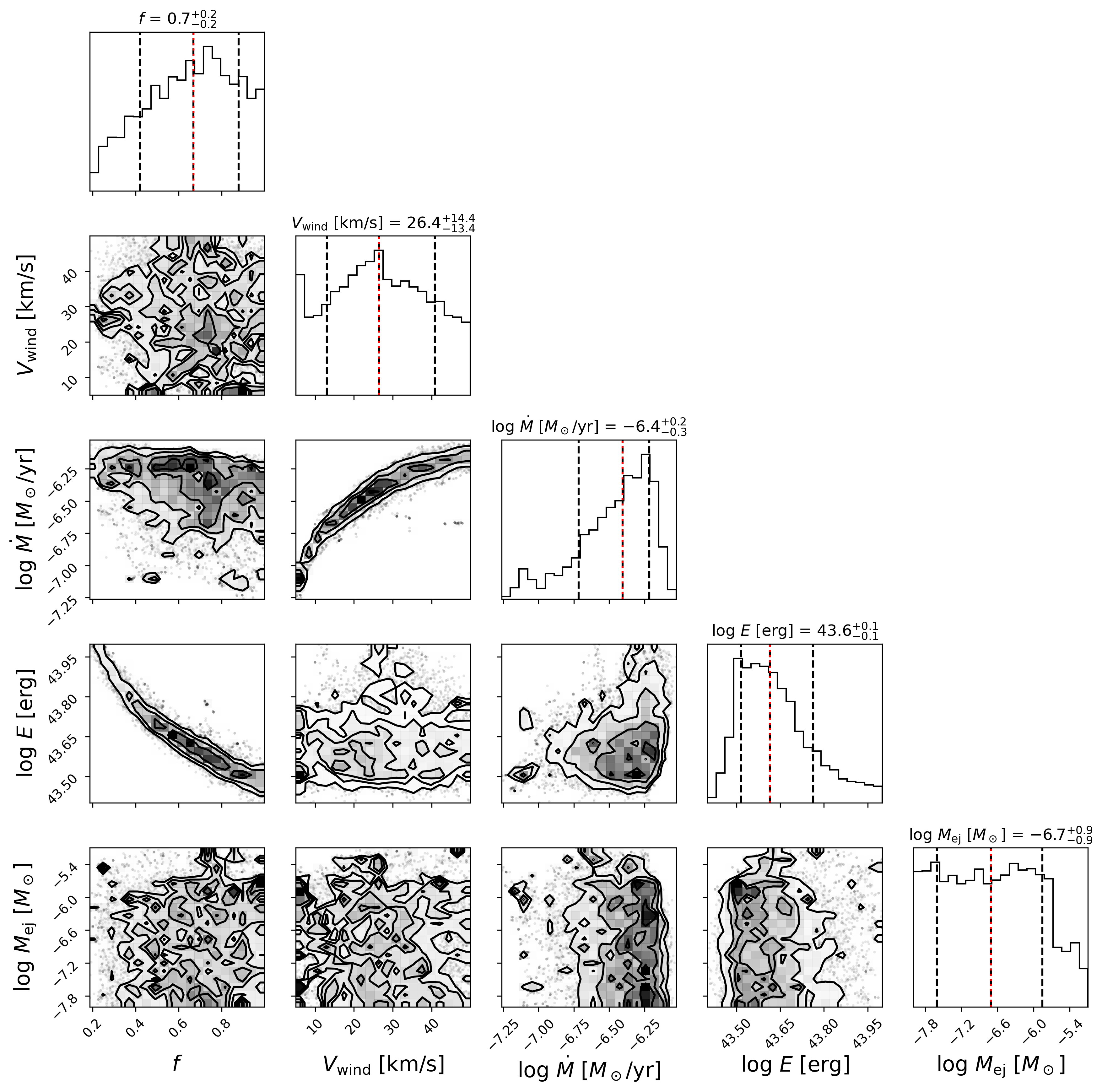} 
  \caption{This is the corner plot of V6598 Sgr, showing the posterior distribution of the five parameters fit with the modified synchrotron emission model. The parameters that we varied are filling factor $f$, wind velocity $V_{\rm{wind}}$, mass-loss rate $\dot M$ energy $E$, ejected mass $M_{\mathrm{ej}}$. This 2D contour plot corresponds to the best-fitting model among the three fits. The correlations between the parameters show that they are coupled and adjust accordingly to maintain the optical evolution.}
  \label{fig:Corner_Plot_6598}
\end{figure*}

However, none of the models gives a satisfactory fit, and the posterior distribution in Figure \ref{fig:Corner_Plot_6598} shows strong degeneracies among the physical parameters. Different combinations of explosion energy, ejecta mass, mass-loss rate, wind velocity, and density-profile shape can produce broadly similar light-curve behaviour. Therefore, the fitted parameters should be interpreted cautiously.

\subsection{V1716 Sco}

\par
V1716 Sco is a classical nova with significant gamma-ray detection. Following the typical classical nova structure, it contains a main-sequence star \citep{2025ApJ...993..232S} as its companion. Therefore, the non-thermal emission from this source indicates internal shocks within the ejecta. The presence of this early radio synchrotron flare is consistent with other gamma-ray-detected novae. From Figure \ref{fig:lightcurves}, we observed that within $\sim 110\,\mathrm{days}$ from the optical eruption, the radio flux rose very steeply and reached its peak, followed by a very steep post-peak decline. This rapid variability at the low frequency $887.5\,\mathrm{MHz}$ makes it a strong synchrotron-dominated source with almost no thermal emission in early times. 

\par
\begin{figure}[htbp]
  \centering
  \includegraphics[width=1\textwidth]{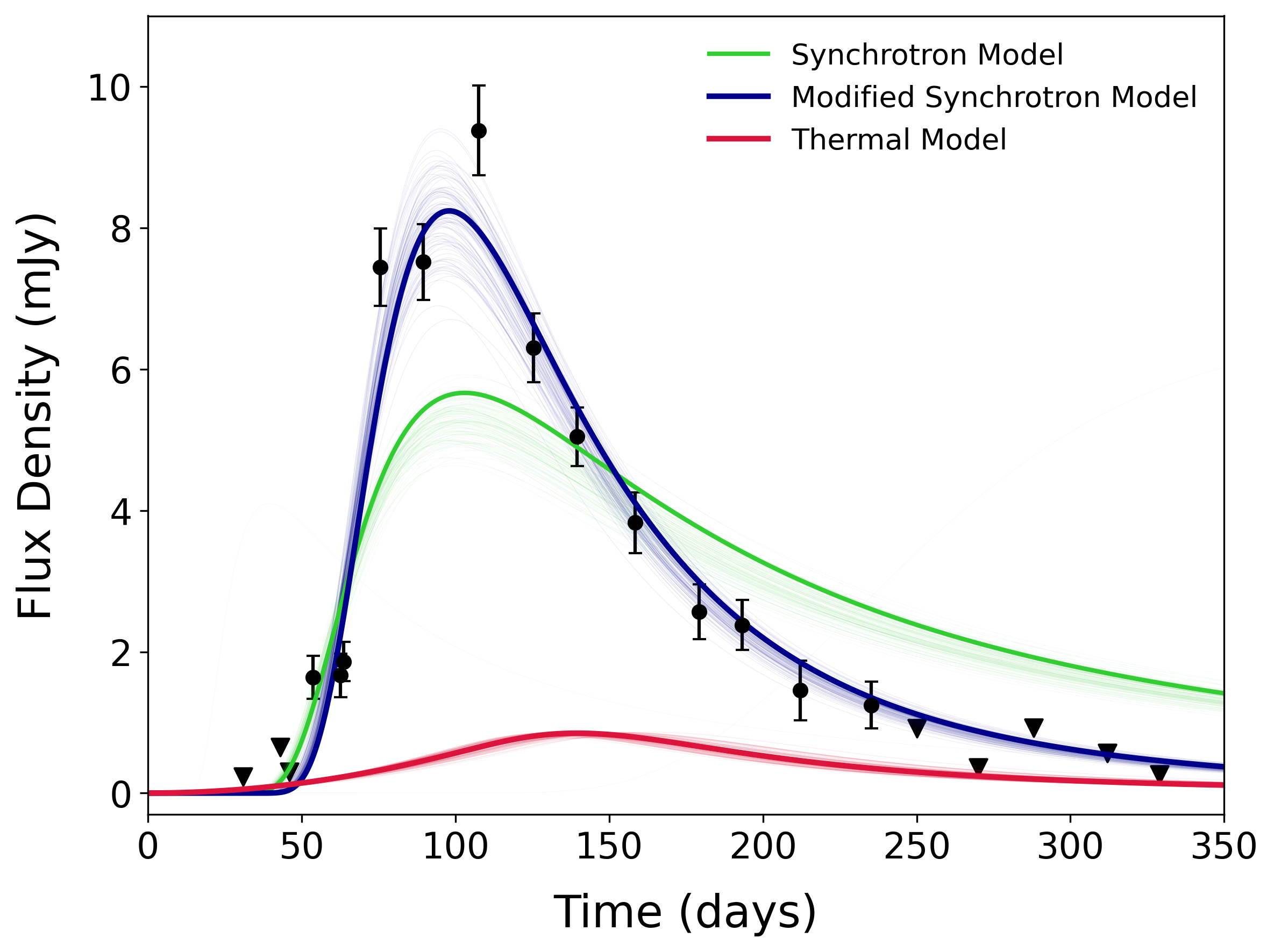} 
  \caption{Radio light curve fitting of V1716 Sco for thermal, non-thermal, and modified non-thermal emission models. Red, green, and blue curves show three different models.  From the MCMC posterior distribution, 100 samples are used to generate a spaghetti plot illustrating the uncertainty ranges, with bold lines indicating the respective best-fit solutions.}
  \label{fig:V1716_ModelFit}
\end{figure}

Figure \ref{fig:V1716_ModelFit} shows the model fits with uncertainties. As expected, the thermal model failed to estimate such a high flux density at a lower frequency, whereas the synchrotron models roughly estimated the observed values. Compared to other classical novae observed at the same frequency, $887.5\,\mathrm{MHz}$, such as V1369 Cen, V5668 Sgr, RR Tel, and YZ Ret \citep{2023PASA...40...25G}, V1716 Sco shows much denser temporal sampling with multiple detections. This suggests that this source is highly luminous and produces very efficient particle acceleration. Since the shock propagates through a complex, non-uniform, steeply graded, dense circumbinary material, a wind-like density profile varying as $R^{-2}$ is not sufficient to describe the evolution. Given that this binary hosts a main-sequence star, it can be assumed that this steep fall after the turnover frequency is due to the very low density of the ambient medium. That is, when the shock propagates through the $R\gg R_0$ region, the density drops so abruptly that there are not enough electrons to be accelerated by the shock-carried energy. This lack of efficiently accelerated electrons causes the flux density to drop very rapidly. Even though none of the models has an acceptable fit based on the $\chi^2$ value, the broken power-law model still provides a better fit and captures the steep decline. We modelled the radio light curves by fixing the slopes to $k=3.1$ for the optically thick phase, $m=3.3$ for the optically thin phase, and $n=2.7$ for the transition between the two. From the fit, it is evident that both the rise ($S_\nu\propto t^{3.1}$) and decay ($S_\nu\propto t^{-3.3}$) are far more rapid than expected for freely expanding, optically thick thermal ejecta, which scales as $S_\nu \propto t^2$.

\par
\begin{figure*}[htbp]
  \centering
  \includegraphics[width=1\textwidth]{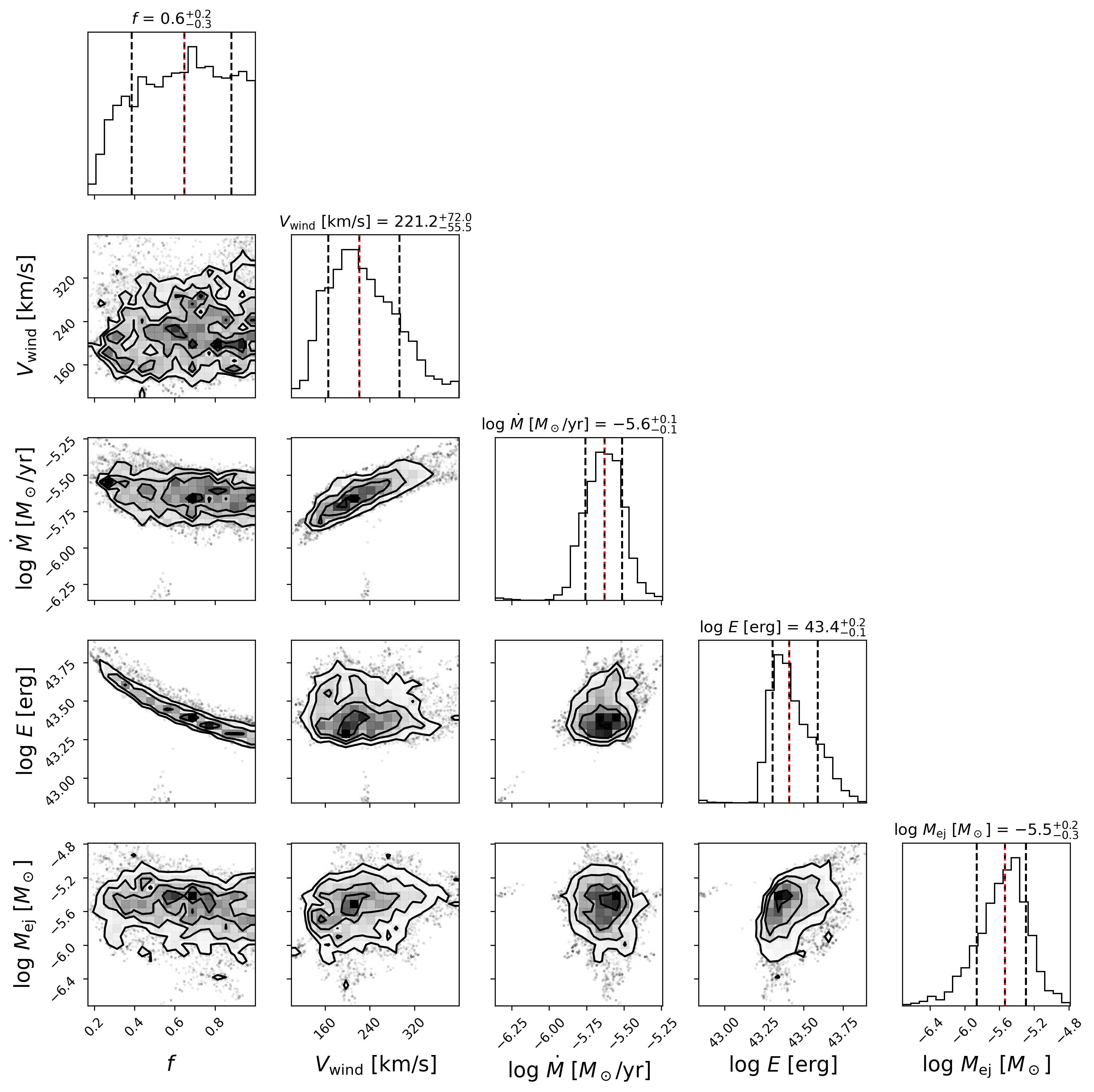} 
  \caption{Posterior distribution of the parameters in the broken power-law synchrotron model fit to V1716 Sco. These parameters are poorly constrained, resulting in a poor fit to the data. Mass loss rate $\dot{M}$ and wind velocity $V_{\mathrm{wind}}$ showed a positive correlation, reflecting the model degeneracy to sustain the observed radio flux, maintaining the $\rho\propto \frac{\dot M}{V_{\mathrm{wind}}}$ density profile. While the Energy and filling factor showed negative correlations, implying that if we make the shock more energetic, we need a smaller filling factor to produce the same radio flux peak.}
  \label{fig:Corner_Plot_1716}
\end{figure*}

The best-fit values for all three models from the MCMC sampling are in Tables \ref{tab:Thermal_Model}, \ref{tab:Synchrotron_Model}, \ref{tab:Mod_Synchrotron_Model}, and Figure \ref{fig:Corner_Plot_1716} is the corner plot of the posterior distribution of our modified synchrotron model fit. The density profile in this model scales as $\rho\propto \dot M\,V^{-1}_{\mathrm{wind}}$, which means that a high mass loss rate needs a higher wind velocity to match the temporal evolution. This explains the parameter degeneracy, contributes to poor constraints, and reduces the quality of the fit. Similarly, from Table \ref{tab: Distance Scaling_nonthermal}, we could compute how the parameters will change if our adopted distance is not true.

 \par
 The 2D corner plot in Figure \ref{fig:Corner_Plot_1716} shows broad but correlated posterior distributions of $f$, $V_{\mathrm{wind}}$, $\dot M$, $E$ and $M_{\mathrm{ej}}$. These degeneracies are expected due to the limited observational constraints. The fractional value $f=0.6$ suggests the partial involvement of the shell in this non-thermal emission. Even though none of the fits was good, the comparison still favours a two-component shock scenario, in which a dense, fast flow collides with a slower, less dense environment, producing non-thermal synchrotron emission that dominates at low frequencies \citep{2016MNRAS.463..394V}.

\subsection{V1723 Sco}

\par
\begin{figure}[htbp]
  \centering
  \includegraphics[width=1\textwidth]{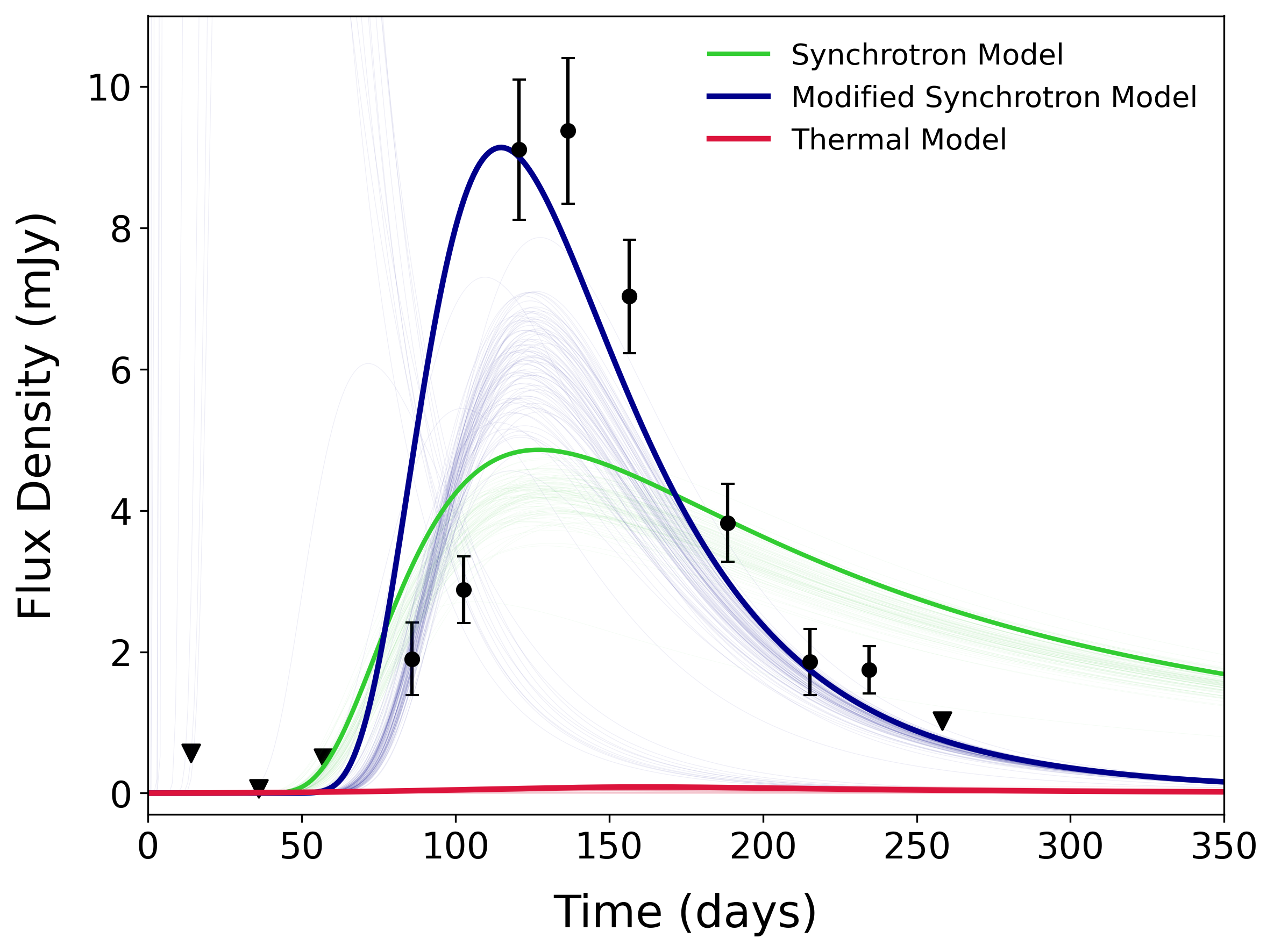} 
  \caption{Radio light curve fitting of V1723 Sco for thermal, non-thermal, and modified non-thermal emission models. Red, green, and blue curves show three different models. From the MCMC posterior distribution, 100 samples are used to generate a spaghetti plot illustrating the uncertainty ranges, with bold lines indicating the respective best-fit solutions.}
  \label{fig:V1723_ModelFit}
\end{figure}

We applied the same three models to this newly observed source, as illustrated in Figure \ref{fig:V1723_ModelFit}. Gamma-ray and X-ray studies indicated this as a very powerful shock-driven non-thermal synchrotron flare with efficient particle acceleration. The flux density rises steeply from $2.8\pm0.3\,\mathrm{mJy}$ to $9.3\pm0.3\,\mathrm{mJy}$. This rapidly evolving synchrotron flare, spanning $\sim 150\,\mathrm{days}$, has no acceptable fit with any of our models. The fitting results are presented in Table \ref{tab:AIC_BIC}. The basic non-thermal model was unable to produce these sharp slopes and failed to account for the short-lived peak. In our broken power-law model, we varied $k$, $m$, and $n$ over a realistic range of $1-4$ to see whether it could produce such steep changes. Later, we set $k=3.1$, $m=3.6$, and $n=3$ for the fitting, but it still failed to match the peak and yielded unsatisfactory $\chi^2$ values. This temporal change $S_\nu\propto t^{-3.6}$ in flux density is much higher than that of a typical thermal-dominated classical nova, decaying as $S_{\nu}\propto t^{-2}$, supporting it as a very bright, luminous, and strong synchrotron emission source with significantly efficient particle acceleration.

\par
The V-band data show some irregularity around the rise and peak epochs of this radio flare. However, we do not interpret this as definitive evidence of correlated optical-radio variability. We note that a recent study of the \textit{Fermi}-LAT gamma-ray observation of V1723 Sco by \citet{2026A&A...705A..19F} supports the presence of shock activity in this system. A possible late-time high-energy activity is mentioned there, which could indicate continued particle acceleration, potentially associated with hot materials heated by a new shock. This shows that V1723 Sco was a shock-active system and not evolving toward a simple spherically symmetric photosphere. Energy injections into the ejecta can also exhibit variability and randomness, which can cause the ejecta to react more violently and generate stronger internal shocks \citep{2021ApJS..257...49C}. Taken together, this supports the idea of an early synchrotron flare arising when the nova is still re-energising or reshaping the entire outflow structure.

\begin{figure*}[htbp]
  \centering
  \includegraphics[width=1\textwidth]{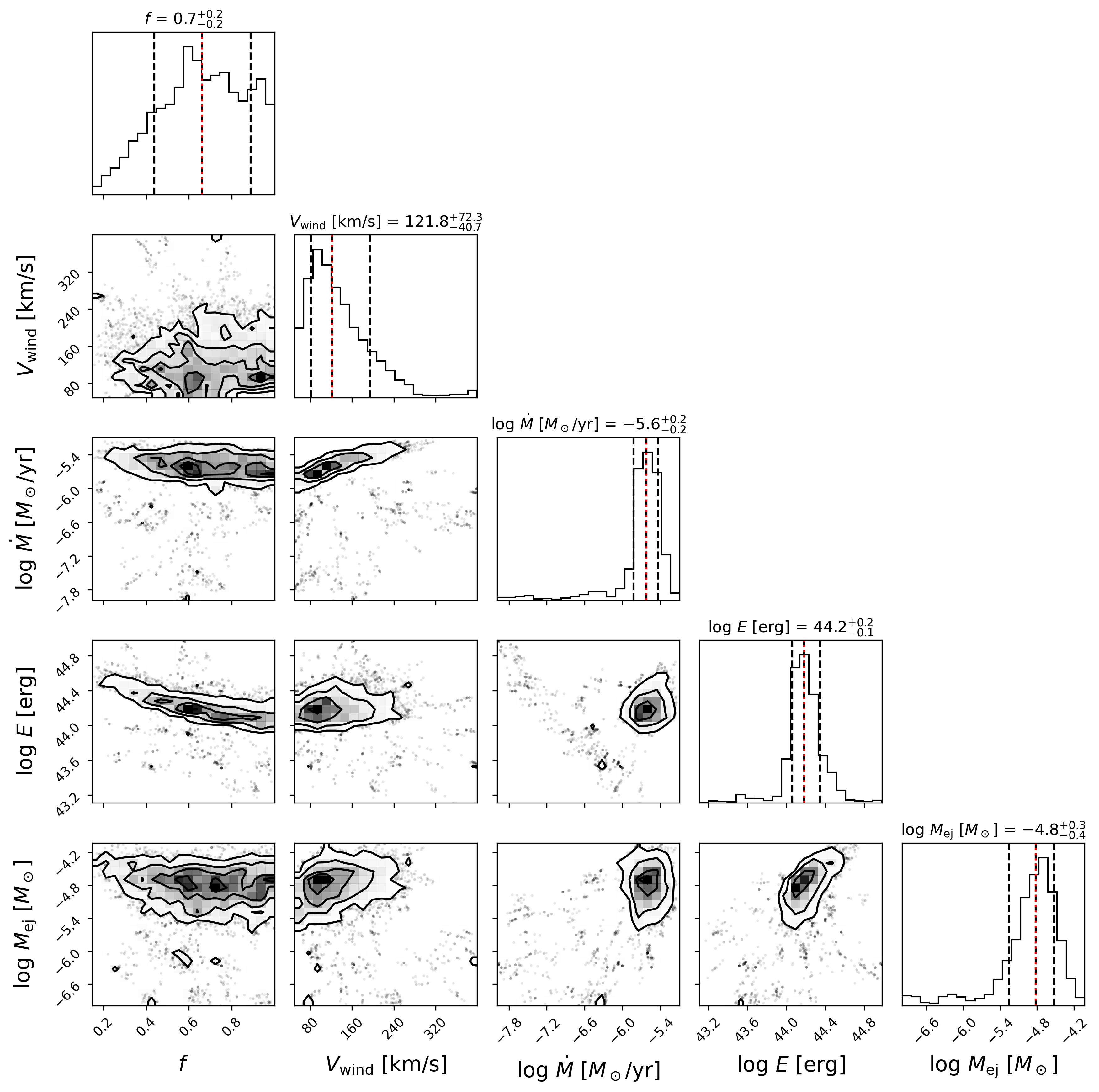} 
  \caption{This is the corner plot showing the posterior distributions from the broken power law synchrotron model fit with the lowest $\chi^2$ value via MCMC analysis of V1723 Sco. Negative correlations in the energy-filling factor contour and a mildly positive correlation in the mass loss-wind velocity contour are observed; other than that, no significant correlations are spotted.}
  \label{fig:1723_Corner_Plot}
\end{figure*}

\begin{table*}[t]
  \centering
  \caption{Best-fit parameters $M_{\mathrm{ej}}$, $v_2$, and $v_1/v_2$ that were directly sampled by the MCMC with the free-free thermal model. We kept the electron temperature $T_{\mathrm{e}}$ fixed at $10^4\,\mathrm{K}$. The inner velocity $v_1$ was not independently varied as a free parameter, but it was computed for each posterior sample by using $v_1/v_2$ with the uncertainty of this derived distribution. The best-fit value of $M_{\mathrm{ej}}$ is mentioned on a $\log$ scale. }
  \label{tab:Thermal_Model}
  \setlength{\tabcolsep}{2pt}
  \renewcommand{\arraystretch}{1.4}

  \begin{tabular*}{\textwidth}{@{\extracolsep{\fill}} l c c c c c c}
    \hline
    \textbf{Source} & $\log M_{\mathrm{ej}}\,(\mathrm{M_{\odot}})$ & $v_2\,(\mathrm{km\,s^{-1}})$ & $v_1/v_2$ & $v_1\,(\mathrm{km\,s^{-1}}) $ & $T_\mathrm{e}\,(\mathrm{K})$ & $d\,(\mathrm{kpc})$ \\
    \hline

     V6598 Sgr & $-3.1\pm0.15$ &  $9800\pm200$  & $0.6\pm0.1$ & $6300\pm1250$ & $10^4$ & 7.6  \\

     V1716 Sco &   $-4.0\pm0.01$  & $9300\pm530$ & \phantom{0}$0.8\pm0.01$ & $7330\pm\phantom{0}440$ & $10^4$  & 3.0 \\

     V1723 Sco & $-4.1\pm0.16$  & $9880\pm350$ & $0.7\pm0.1$ & \phantom{0}$6900\pm1000$ &$10^4$ & 8.0 \\
    \hline
  \end{tabular*}
\end{table*}

\begin{table*}[t]
  \centering
  \caption{Best-fit parameters $f$, $V_{wind}$, $\dot M$, $E$, and $M_{ej}$ that were directly sampled by the MCMC with the synchrotron emission model fit. We kept a fixed value of $p=2.5$, and $\epsilon_{\mathrm{e}}=\epsilon_{\mathrm{B}}=0.01$ and distance $d$ according to the source. The best-fit values of $\dot M$, $E$, $M_{\mathrm{ej}}$ parameters are in $\log$ scale.}
  \label{tab:Synchrotron_Model}
  \setlength{\tabcolsep}{2pt}
  \renewcommand{\arraystretch}{1.4}

  \begin{tabular*}{\textwidth}{@{\extracolsep{\fill}} l c c c c c c c c c}
    \hline
    \textbf{Source} & $f$ & $V_{\mathrm{wind}}\,(\mathrm{km\,s^{-1}})$ &  $\log \dot{M}\,(\mathrm{{M}_{\odot}\,yr^{-1}})$ & $\log E\,(\mathrm{erg})$ & $\log M_{\mathrm{ej}}\,(\mathrm{{M}_{\odot}})$ & $p$ & $\epsilon_\mathrm{e}$ & $\epsilon_\mathrm{B}$ & $d\,\mathrm{(kpc)}$  \\
    \hline

    V6598 Sgr  & $0.6\pm0.2$ &  \phantom{0}$46\pm\phantom{0}32$ & $-6.3\pm0.4$ & $43.5\pm0.2$ & $-5.8\pm0.6$ & 2.5 & 0.01 & 0.01 &  7.6   \\

    V1716 Sco &  $0.7\pm0.2$  & $213\pm110$ & $-5.9\pm0.2$ & $43.0\pm0.1$ & $-5.9\pm0.3$ & 2.5 & 0.01 & 0.01 &  3.0  \\

    V1723 Sco &  $0.7\pm0.2$  & $223\pm\phantom{0}90$ &  $-5.5\pm0.2$ & $43.6\pm0.2$ &  $-5.6\pm0.5$ &  2.5 & 0.01 & 0.01 & 8.0 \\

    \hline
  \end{tabular*}
\end{table*}

\begin{table*}[t]
  \centering
  \caption{In this Table the best-fit parameters $f$, $V_{\mathrm{wind}}$, $\dot M$, $E$, and $M_{\mathrm{ej}}$ were listed from the modified synchrotron emission model fits. $\dot M$, $E$, $M_{\mathrm{ej}}$ are listed in $\log$ scale. Here, we kept $p$, $\epsilon_{\mathrm{e}}$, and $\epsilon_{\mathrm{B}}$ at values similar to those of the previously fitted standard synchrotron model. However, from the broken-power law radial function in $\rho$ profile, we now fixed three new indices $k$ (controls inner slope), $m$ (controls outer slope), $n$ (transition rate) with different values based on the rise and decay slopes. We assigned these powers individually to the source, as each showed a different temporal behaviour. For example, at early times, V6598 Sgr had a typical value of $t^{-2}$ in its slope, whereas V1716 Sco and V1723 Sco varied much more steeply at $\approx t^{-3.1}$. We can clearly observe this rapid variability in their light curves before even applying any model.}
  \label{tab:Mod_Synchrotron_Model}
  \setlength{\tabcolsep}{2pt}
  \renewcommand{\arraystretch}{1.4}

  \begin{tabular*}{\textwidth}{@{\extracolsep{\fill}} l c c c c c c c c c c c c }
    \hline
    \textbf{Source} & $f$ & $V_\mathrm{wind}\,(\mathrm{km\ s^{-1}})$ &  $\log\dot {M}\,(\mathrm{M_\odot\,yr^{-1}})$ & $\log E\,(\mathrm{erg})$ & $\log M_\mathrm{ej}\,(\mathrm{M_\odot})$ & $p$ & $\epsilon_\mathrm{e}$ & $\epsilon_\mathrm{B}$ & $d\,(\mathrm{kpc})$ & $k$ & $m$ & $n$  \\
    \hline

    V6598 Sgr  & $0.7\pm0.2$  &   $26\pm14$ & $-6.4\pm0.3$ & $43.6\pm0.1$ & $-6.7\pm0.9$  & 2.5 & 0.01 & 0.01 &  7.6 & 2.4 & 2.5 & 0.7 \\

    V1716 Sco &  $0.6\pm0.3$  & $221\pm64$\phantom{0} & $-5.6\pm0.1$ &  $43.4\pm0.2$ & $-5.5\pm0.3$  & 2.5 & 0.01 & 0.01 &  3.0  & 3.1 & 3.3 & 2.7 \\

    V1723 Sco &  $0.7\pm0.2$ & $122\pm56$\phantom{0} & $-5.6\pm0.2$ & $44.2\pm0.2$ &  $-4.8\pm0.4$ &  2.5 & 0.01 & 0.01 & 8.0 & 3.1 & 3.6 & 3.0 \\

    \hline
  \end{tabular*}
\end{table*}

\begin{table*}
\caption{Model comparison statistics for each nova. Lower $\chi^2_\nu$, AIC, and BIC values indicate a better fit. Based on their calculated values, the preferred model is mentioned. V1723 Sco had no acceptable fit with any of the models. Therefore, we are unable to identify the preferred model for this source.}
\label{tab:AIC_BIC}
\begin{tabular*}{\textwidth}{@{\extracolsep{\fill}}l l c c c c}
\hline
Source & Model & AIC & BIC & Reduced $\chi^2$ & Preferred emission model\\

\hline
V6598~Sgr & Thermal     & $334$ & $335$ & $65.7$ &  \\ 
          & Synchrotron & \phantom{0}$37$ & \phantom{0}$37$ & \phantom{0}$8.9$ & Poor fit to the modified Synchrotron model\\ 
          & Modified     & \phantom{0}$18$ & \phantom{0}$19$ &  \phantom{0}$2.9$ &  \\  

\hline
V1716~Sco & Thermal     & $510$ & $512$ & $50.4$  & \\
          & Synchrotron & $145$ & $147$ & $73.4$ & Poor fit to the modified Synchrotron model \\
          & Modified      & \phantom{0}$40$ & \phantom{0}$42$ & \phantom{0}$3.7$ &  \\

\hline
V1723~Sco & Thermal     & $380$ & $380$ & $74.7$ & \\
          & Synchrotron & $189$ & $190$ & $59.0$ & No model has an acceptable fit \\
          & Modified    & $123$ & $124$ & $37.8$ &  \\
\hline
\end{tabular*}
\end{table*}

\begin{figure}[htbp]
  \centering
  \includegraphics[width=\textwidth]{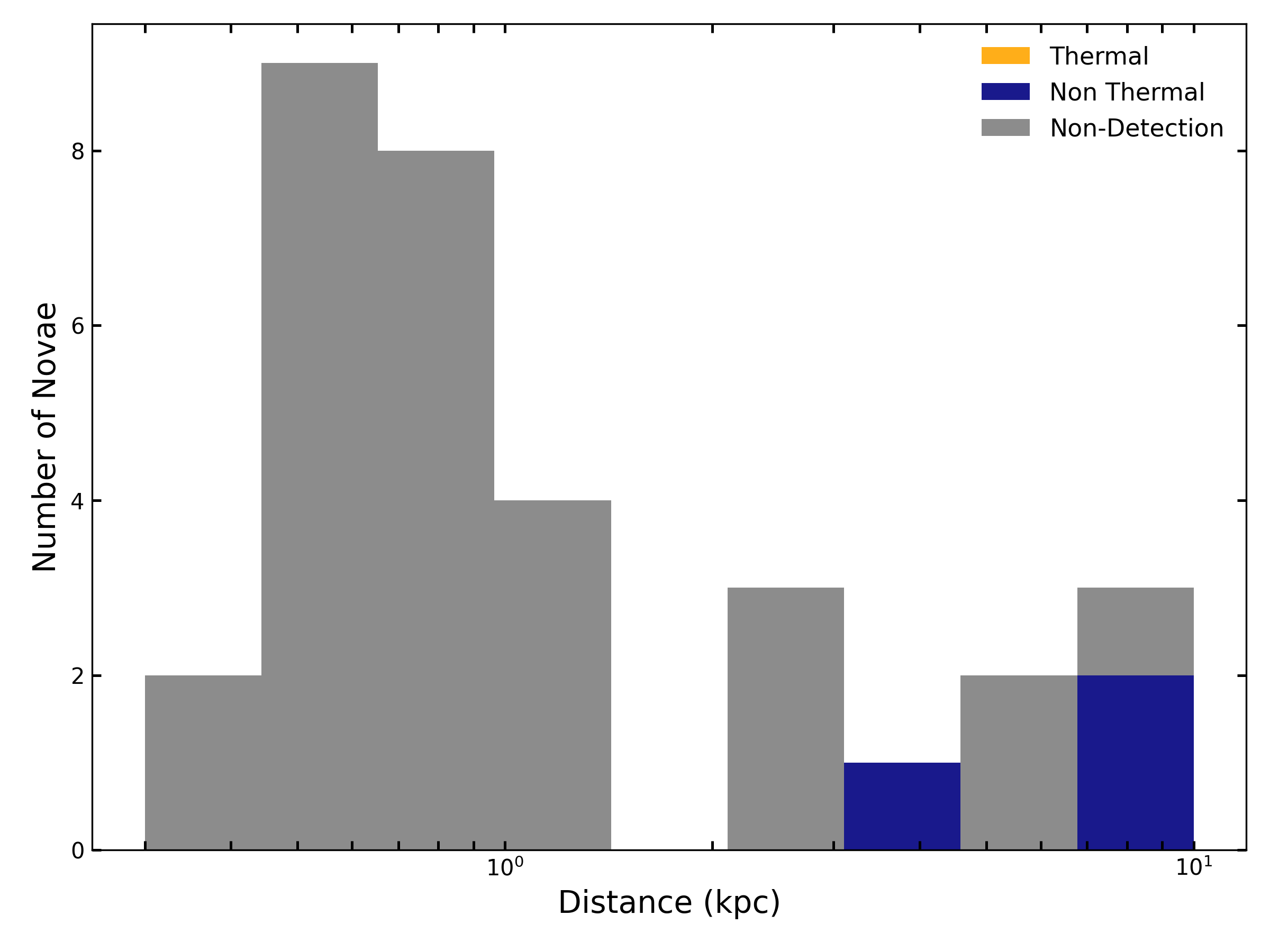} 
  \caption{Distribution of classical novae as a function of distance (kpc), separated by emission mechanisms. The histogram uses logarithmically spaced distance bins to provide equal-interval spacing. Most of the novae within $1\,\mathrm{kpc}$ show either non-detections or thermal radiation, while at large distances ($\geq4\,\mathrm{kpc}$) the sample is dominated by the brighter, luminous non-thermal novae detectable in the ASKAP at $887.5\,\mathrm{MHz}$.}
  \label{fig:Distribution_CNe}
\end{figure}

\begin{figure}[htbp]
  \centering
  \includegraphics[width=\textwidth]{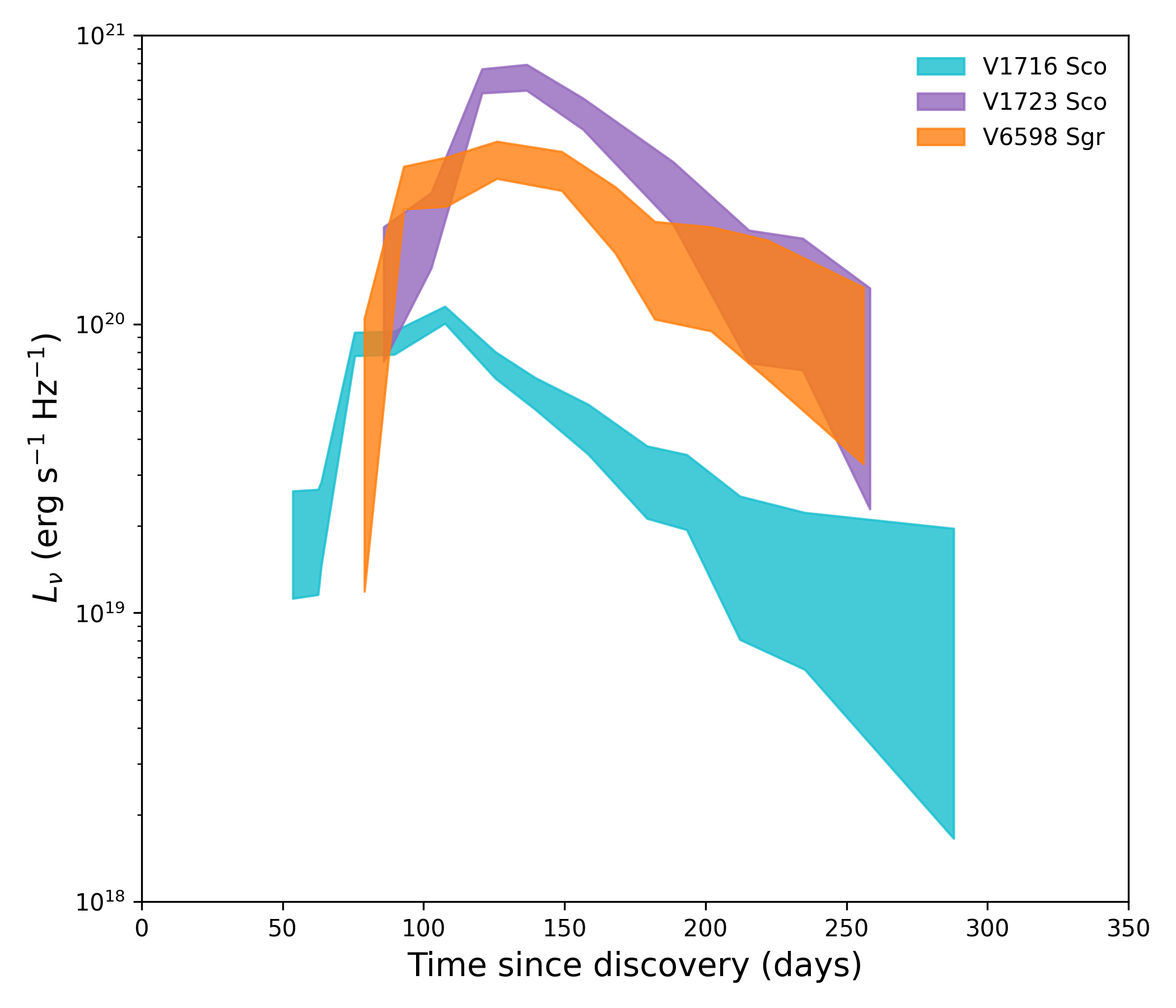} 
  \caption{Radio luminosity is plotted as observed at $887.5\,\mathrm{MHz}$ assuming the nominal distance for each source; the width of the shaded region denotes uncertainties on flux density.}
  \label{fig:Spectral_Luminosity}
\end{figure}

\begin{figure}[htbp]
  \centering
  \includegraphics[width=\textwidth]{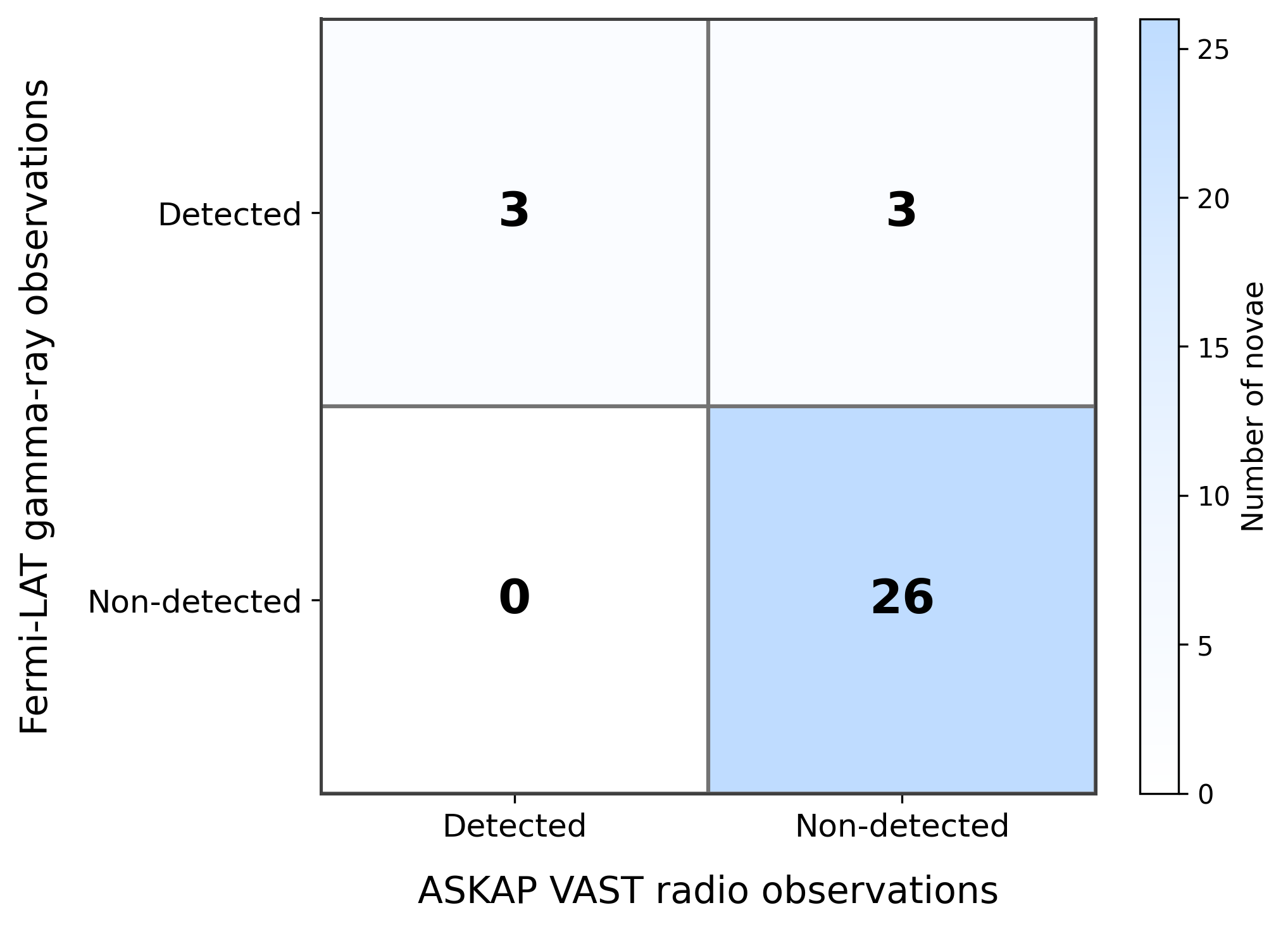} 
  \caption{Comparison of \textit{Fermi}-LAT $\gamma$-ray and ASKAP VAST radio detections of novae, to assess whether radio detections are preferentially associated with prior gamma-ray detections. Each cell gives the number of novae in the corresponding detection category. The colour intensity scales with the number of sources per cell, with darker shading indicating a higher count. The estimated $p\simeq0.004$ value from the binomial confidence interval shows that there is only about a $0.4\%$ chance of getting a distribution like this, where all 3 ASKAP radio-detected novae fall within 6 \textit{Fermi}-LAT gamma-ray-detected novae.}
  \label{fig:LAT_ASKAP}
\end{figure}

\section{Discussion}
\label{sec:discussion}

\par
 We modelled the light curves of these novae using two emission mechanisms commonly observed in classical novae: thermal free-free emission \citep{1979AJ.....84.1619H} and non-thermal synchrotron emission \citep{1998ApJ...499..810C, 2023MNRAS.523.1661N}. Again, in contrast to previous works, where identification of non-thermal emission was based on spectral information \citep{2024MNRAS.534.1227M,2016MNRAS.457..887W, 2016MNRAS.460.2687W, 2021MNRAS.501.1394N} or brightness temperature measurements \citep{2021ApJS..257...49C}, here we attempted to identify the dominant emission mechanism based on the light curve shape. We adopted a broken power-law in the synchrotron emission model to incorporate non-uniform shock-evolution dynamics and referred to it as the modified synchrotron emission model. For all three novae, the free-free thermal model under-predicts the flux density and was not able to describe the observed light curve. The synchrotron model over-predicted the flux and failed to reproduce the temporal evolution. For these sources, the modified synchrotron model provides a better fit than the thermal or standard synchrotron models. Although none of the models has a formally acceptable fit, this model with a broken power-law density profile yielded the lowest $\chi^2$ among the three. This early synchrotron flux depends primarily on the density of the circumbinary material and on how the faster ejecta interacts with the lower, less dense material, producing internal shock waves driven by the sharp velocity and density gradient. All these systems have an MS as their companion \citep{2025ApJ...993..232S}, which implies a tenuous circumbinary medium. Due to the absence of a dense circumstellar medium, the duration of this early flare is very rapid and short-lived. Therefore, the steep post-decay observed in these sources is likely because, as the wave propagates into a much lower-density medium, it encounters no electrons in front of it to accelerate. In this work, we adopted a simple spherically symmetric model, whereas nova ejecta are far more complex in density structure and morphologies. Radio observations \citep{2021ApJS..257...49C} of systems such as V745 Sco \citep{2024MNRAS.534.1227M}, V445 Pup \citep{2021MNRAS.501.1394N}, V1723 Aql \citep{2014ASPC..490..339W} revealed complex mass loss behaviour, including rapid variability and multiple peaks in their radio light curves. Therefore, the flux density evolution of these systems cannot be reproduced by a single thermal or non-thermal emission model. Future multi-frequency radio observations combined with improved distance constraints and more physically realistic shock models will be essential for better constraining emission dynamics and particle acceleration processes in these systems.

\par
In Figure \ref{fig:Distribution_CNe}, we present the distribution of classical novae as a function of distance. From the literature survey, we find that most novae within $1-2\,\mathrm{kpc}$ remained undetected by ASKAP VAST. This may be partly explained by our observing frequency of $887.5\,\mathrm{MHz}$. At this low frequency, many nearby novae are not detected because they remain optically thick, with the radio emission suppressed by free-free or synchrotron self-absorption. Therefore, we see that classical novae detected by ASKAP VAST within approximately 100 to 150 days after eruption exhibit relatively early, low-frequency radio emission, compared with many thermally dominated nova light curves, which can peak on much longer timescales. These early emissions suggest that these sources are intrinsically radio luminous, have rapidly decreasing radio-optical depths, and contain shock-powered non-thermal emission. Therefore, we examined spectral luminosity to compare the intrinsic radio power of each source. Figure \ref{fig:Spectral_Luminosity} shows all three novae are very luminous ($\geq10^{20}\,\mathrm{erg}\,\mathrm{s}^{-1}\,\mathrm{Hz}^{-1}$). Such high luminosity is difficult to explain with purely thermal emission at late times. Previously studied classical novae with strong synchrotron emission, such as V392 Per, V1723 Aql, V357 Mus, V959 Mon, V838 Her, and V5589 Sgr \citep{2021ApJS..257...49C} had similar spectral luminosity.

\par
In Figure \ref{fig:LAT_ASKAP} we compare \textit{Fermi}-LAT and ASKAP VAST detections to show that among the full ASKAP-footprint sample of 32 optically discovered classical novae from 2021 September to 2025 September, only six have reported gamma-ray emission. Notably, all three ASKAP VAST-detected novae in this work fall within this population. The radio detection fraction is $3/6=50\%$, with a $95\%$ binomial confidence interval of approximately $12\%$ to $88\%$. None of the 26 novae without \textit{Fermi}-LAT detections were detected by ASKAP, corresponding to a radio detection fraction of $0/26=0\%$ with a $95\%$ upper confidence limit of approximately $13\%$. From this result, we estimated the $p$ value as $p\simeq0.004$ from the $2\times2$ contingency table in Figure \ref{fig:LAT_ASKAP}, showing that the radio detection rate by ASKAP is significantly higher among gamma-ray-detected novae. The fact that ASKAP detected a subset that lies entirely within this LAT-detected population suggests that these radio observations may preferentially detect novae with strong shock activity or high intrinsic radio luminosities. This may suggest that gamma-ray emission is an early indicator of shock waves and particle acceleration processes, while later radio synchrotron emission reflects the same shock activity. From the observed radio evolution, we have seen that the timescale for the ejecta to become sufficiently transparent is approximately 100--150 days after the optical eruption. However, because only three of the six LAT-detected novae are detected by ASKAP, LAT detection is not sufficient to guarantee a radio detection. Sensitivity, cadence, distance, and intrinsic differences in the ejecta or circumbinary environment likely also affect whether a nova is detectable at $887.5\,\mathrm{MHz}$.

\par
This shows how the radio evolution observed in classical novae can be affected by ejecta geometry, shock-wave properties, the density distribution of the circumstellar environment, and the emission mechanisms. It also highlights the limitations of simplistic models and the need for more complex, physically motivated models in subsequent studies.

\section{Conclusion}
\label{sec:conclusion}

In this paper, we fitted the radio light curves of three classical novae observed at $887.5\,\mathrm{MHz}$ with a two-week cadence. The model-fitting results can be summarised as follows:

\begin{itemize}

    \item \textbf{V6598 Sgr} is best described by modified synchrotron emission rather than the other two emission models. It peaks at $5.3\pm0.26\,\mathrm{mJy}$ and is consistent with a shock-driven non-thermal synchrotron source.
    \item \textbf{V1716 Sco} also exhibits an early non-thermal synchrotron flare. A broken power-law model provides a better fit than a single power-law model, indicating rapid evolution and efficient shock-driven particle acceleration.
    \item \textbf{V1723 Sco} is not well described by either model, though the broken power-law model captures the steep temporal evolution better than the single power-law model, suggesting an early non-thermal synchrotron flare with a much more complex ejecta structure.  
\end{itemize}

\citet{2023PASA...40...25G} previously reported radio detections of four classical novae, where  VAST data had limited temporal sampling, which they augmented with dedicated follow-up observations at higher frequencies. In contrast, our work here uses only the ASKAP VAST survey of regular, homogeneous monitoring at $887.5\,\mathrm{MHz}$, enabling a well-sampled low-frequency light curve. This may then provide a template for the type of constraints that are feasible with future radio surveys, such as the Square Kilometre Array \citep{2009IEEEP..97.1482D}, where follow-up observations of large numbers of sources are not feasible.

\par
This study highlights ASKAP's role in identifying intrinsically bright classical novae at low radio frequencies. Especially when the typical thermal emission mechanism shapes the radio light curve and constrains the parameter, this early synchrotron flare at such a low frequency probes an unexplored regime in the nova evolutionary stage. We were able to model the light curves and analyse their rapid variability due to the dynamical evolution of the shock. This shows ASKAP's capabilities to constrain shock energetics, ejecta-circumbinary material interactions, non-uniform ambient density, and a more complicatedly distributed emitting region. In the future, with multi-frequency observations and improved distance constraints, we will provide spectral evolution, a strong and direct diagnostic for identifying the emission mechanism. Therefore, we need unbiased continuous monitoring to obtain higher-quality data for more precise parameter estimation and to improve our understanding of the physics behind these star systems.

\begin{acknowledgement}
We thank an anonymous referee for helpful comments.
This work was done as part of the ASKAP Variables and Slow Transients (VAST) collaboration \citep{2013PASA...30....6M}. DK was supported by NSF grant.
AST-2511757.  This research was supported by the Sydney Informatics Hub, a Core Research Facility of the University of Sydney. Parts of this research were supported by the Australian Research Council Centre of Excellence for Gravitational Wave Discovery (OzGrav), project number CE230100016. This scientific work uses data obtained from Inyarrimanha Ilgari Bundara, the CSIRO Murchison Radio-astronomy Observatory. We acknowledge the Wajarri Yamaji People as the Traditional Owners and native title holders of the Observatory site. CSIRO's ASKAP radio telescope is part of the Australia Telescope National Facility (https://ror.org/05qajvd42). Operation of ASKAP is funded by the Australian Government with support from the National Collaborative Research Infrastructure Strategy. ASKAP uses the resources of the Pawsey Supercomputing Research Centre. Establishment of ASKAP, Inyarrimanha Ilgari Bundara, the CSIRO Murchison Radio-astronomy Observatory, and the Pawsey Supercomputing Research Centre are initiatives of the Australian Government, with support from the Government of Western Australia and the Science and Industry Endowment Fund.

\end{acknowledgement}
\clearpage

\bibliography{references}


\end{document}